\documentclass[twocolumn, twocolappendix]{aastex631}

\newcommand\sersic{S\'ersic}
\newcommand\gamornet{G\textsc{a}M\textsc{or}N\textsc{et}}
\newcommand\gampen{GaMPEN}
\newcommand\gb{\textit{g}}
\newcommand\rb{\textit{r}}
\newcommand\ib{\textit{i}}
\newcommand\zb{\textit{z}}
\newcommand\yb{\textit{y}}
\newcommand\gbin{\textit{low}}
\newcommand\rbin{\textit{mid}}
\newcommand\ibin{\textit{high}}
\newcommand\zbin{\textit{extra}}
\newcommand\ybin{\textit{extreme}}
\newcommand\gBin{\textit{Low}}
\newcommand\rBin{\textit{Mid}}
\newcommand\iBin{\textit{High}}
\newcommand\zBin{\textit{Extra}}
\newcommand\yBin{\textit{Extreme}}
\newcommand\sig{\textrm{sig}}

\shorttitle{Machine Learning Framework to Study AGN Host Galaxy Morphological Parameters}
\shortauthors{Tian et al.}

\usepackage{amsmath}
\usepackage{amssymb}
\usepackage{hyperref}
\usepackage{CJK}
\usepackage{soul}

\let\OLDthebibliography\thebibliography
\renewcommand\thebibliography[1]{
  \OLDthebibliography{#1}
  \setlength{\parskip}{0pt}
  \setlength{\itemsep}{0pt plus 0.3ex}
}

\begin{document}
\begin{CJK*}{UTF8}{gbsn}

\title{Automatic Machine Learning Framework to Study Morphological Parameters of AGN Host Galaxies within $z<1.4$ in the Hyper Supreme-Cam Wide Survey}

\author[0000-0003-4056-7071]{Chuan Tian (田川)}
\affiliation{Department of Physics, Yale University, New Haven, CT, USA}
\email{chuan.tian@yale.edu; ufsccosmos@gmail.com}

\author[0000-0002-0745-9792]{C. Megan Urry}
\affiliation{Department of Physics, Yale University, New Haven, CT, USA}
\affiliation{Department of Astronomy, Yale University, New Haven, CT, USA}
\affiliation{Yale Center for Astronomy and Astrophysics, Yale University, New Haven, CT, USA}

\author[0000-0002-2525-9647]{Aritra Ghosh}
\altaffiliation{LSST-DA Catalyst Fellow}
\affil{DiRAC Institute and the Department of Astronomy, University of Washington, Seattle, WA, USA}
\affil{Department of Astronomy, Yale University, New Haven, CT, USA}

\author[0000-0002-6766-5942]{Daisuke Nagai}
\affil{Department of Physics, Yale University, New Haven, CT, USA}
\affiliation{Department of Astronomy, Yale University, New Haven, CT, USA}
\affiliation{Yale Center for Astronomy and Astrophysics, Yale University, New Haven, CT, USA}

\author[0000-0001-8211-3807]{Tonima T. Ananna}
\affiliation{Department of Physics and Astronomy, Wayne State University, Wilder Hall, 17 Fayerweather Hill Road, Hanover, NH, USA}

\author[0000-0003-2284-8603]{Meredith C. Powell}
\affil{Leibniz-Institut fur Astrophysik Potsdam (AIP), An der Sternwarte 16, D-14482 Potsdam, Germany}

\author[0000-0002-5504-8752]{Connor Auge}
\affiliation{
Institute for Astronomy, University of Hawaii, Honolulu, HI, USA}

\author[0000-0003-1164-0268]{Aayush Mishra}
\affil{Department of Physics, Indian Institute of Science Education and Research, Bhopal, India}

\author[0000-0002-1233-9998]{David B. Sanders}
\affiliation{
Institute for Astronomy, University of Hawaii, Honolulu, HI, USA}

\author[0000-0002-1697-186X]{Nico Cappelluti}
\affiliation{Department of Physics, University of Miami, Coral Gables, FL, USA}
\affiliation{INAF - Osservatorio di Astrofisica e Scienza dello Spazio di Bologna, Bologna, Italy}

\author[0000-0001-5464-0888]{Kevin Schawinski}
\affiliation{Modulos AG, Technoparkstr. 1, CH-8005, Zurich, Switzerland}

\begin{abstract}

We present a composite machine learning framework to estimate posterior probability distributions of bulge-to-total light ratio, half-light radius, and flux for Active Galactic Nucleus (AGN) host galaxies within $z<1.4$ and $m<23$ in the Hyper Supreme-Cam Wide survey.
We divide the data into five redshift bins:
\gbin{} ($0<z<0.25$), \rbin{} ($0.25<z<0.5$), \ibin{} ($0.5<z<0.9$), \zbin{} ($0.9<z<1.1$) and \ybin{} ($1.1<z<1.4$), and train our models independently in each bin.
We use PSFGAN to decompose the AGN point source light from its host galaxy, and invoke the Galaxy Morphology Posterior Estimation Network (GaMPEN) to estimate morphological parameters of the recovered host galaxy.
We first trained our models on simulated data, and then fine-tuned our algorithm via transfer learning using labeled real data.
To create training labels for transfer learning, we used GALFIT to fit $\sim 20,000$ real HSC galaxies in each redshift bin.
We comprehensively examined that the predicted values from our final models agree well with the GALFIT values for the vast majority of cases.
Our PSFGAN + \gampen{} framework runs at least three orders of magnitude faster than traditional light-profile fitting methods, and can be easily retrained for other morphological parameters or on other datasets with diverse ranges of resolutions, seeing conditions, and signal-to-noise ratios, making it an ideal tool for analyzing AGN host galaxies from large surveys coming soon from the Rubin-LSST, Euclid, and Roman telescopes.
\end{abstract}


\section{Introduction} \label{sec:intro}
For a long time, AGN activity has been thought to be connected to the evolution of its host galaxy.
Energy generated by an active central supermassive black hole (SMBH) can have a profound effect on suppressing new star formation \citep[e.g.][]{2005Natur.433..604D, 2006MNRAS.365...11C, 10.1111/j.1745-3933.2006.00234.x, 2007ApJ...654..731H, 10.1111/j.1745-3933.2008.00430.x, 2012ARA&A..50..455F, 2014ARA&A..52..589H, 2017NatAs...1E.165H}, which in turn regulates the growth of the SMBH itself \citep{2014ARA&A..52..589H, 2017NatAs...1E.165H}.
This co-evolution picture leads to another question---how are AGNs, especially the most luminous ones, triggered?
Contemporary theories suggest that highly luminous AGNs can be triggered during major merger events \citep[e.g.][]{1988ApJ...325...74S, 2005ApJ...630..705H, 2006ApJS..163....1H, Treister_2012, Draper_2012, 10.1093/mnras/stu1736, 2014MNRAS.440..889S, 2017MNRAS.468.1273R, Glikman_2023}.
A useful proxy for determining whether an AGN has been
triggered by a major merger is the host galaxy morphology
\citep[e.g.][]{2008ApJ...683..597S, 2008MNRAS.391.1137L, 2013MNRAS.433.2986W}.
Therefore, studying the morphological properties of AGN host galaxies can provide important clues on the perplexing question of AGN triggering.

To investigate AGN host galaxy morphology, there are two key points to consider: 
(1) A quantitative, rather than qualitative, study of AGN host galaxy morphology is preferable, in order to provide robust constraints for models \citep[e.g.][]{2014MNRAS.440..889S}. 
(2) AGN activity evolves substantially with redshift \citep{1998ApJ...498..106M, 2003ApJ...582..559V, 2009ApJ...690...20S, doi:10.1146/annurev-astro-081811-125615}, so AGN and host galaxy properties must be ascertained over a broad range of redshift. 
In this paper, we explore AGN host galaxies from the local universe up to $z\sim1.4$, quantitatively analyzing morphological parameters using images from large ground-based surveys.

In recent years, machine learning (ML) techniques have shown remarkable promise in studying galaxy morphology at scale \citep[e.g.][]{2015MNRAS.450.1441D, 2015ApJS..221....8H, 2018MNRAS.475..894T, 2018MNRAS.478.5410D, 2018MNRAS.476.3661D, 2020A&C....3000334B, 2020MNRAS.491.1554W,
2020ApJS..248...20H, 2021MNRAS.506.1927V, 2021MNRAS.503.4446C, 2021MNRAS.507.4425C, 2022MNRAS.511.3330T}. 
Traditional 2D light profile fitting methods, such as GIM2D \citep{1998ASPC..145..108S}, GALFIT \citep{2002AJ....124..266P} and GALAPAGOS \citep{2012MNRAS.422..449B}, 
present a challenge in the face of the data volume expected from next-generation surveys, 
as do visual classifications by trained experts.
In our previous work \citep{2023ApJ...944..124T}, we utilized \gamornet 
 \citep{2020ApJ...895..112G}, a Convolutional Neural Network (CNN; \citealp{726791, Imagenet}) to qualitatively classify AGN host galaxies as disk-dominated, bulge-dominated, or indeterminate, achieving a classification precision of $\sim$ 80\%-95\% across various redshift bins.

In this work, we present and test a composite ML framework to investigate AGN host morphologies; we will use it to quantitatively study morphological parameters of real AGN host galaxies in a future paper.
We use a modified (see Section\,\ref{sec:intro_psfgan} and Figure\,\ref{fg:psfgan_arch} for more details) implementation of PSFGAN \citep{2018MNRAS.477.2513S}, a Generative Adversarial Network (GAN; \citealp{goodfellow2014generative}), to remove the AGN point source that often dominates over the host galaxy.
Unlike 2D light profile fitting methods, PSFGAN does not require explicit knowledge of the point spread function (PSF), and it runs orders of magnitude faster during inference to achieve similar results.
We then study the decomposed host galaxy light using \gampen{} \citep{2022ApJ...935..138G}, the successor of \gamornet, a novel ML framework that is capable of estimating Bayesian posterior distributions for the bulge-to-total light ratio ($L_B/L_T$), the half-light radius ($R_e$), and the total flux ($F$) of the (PSFGAN-recovered) host galaxy.
\gampen{} does this by using the Monte Carlo Dropout technique \citep{gal_2016}, while each input galaxy image is fed to hundreds of slightly different CNNs\footnote{This is done by randomly dropping out some neurons during inference---resulting in hundreds of slightly varying models.}, and their outputs are then aggregated to make probability distributions (see Section~\ref{sec:overv_appro} for more details).
Another unique feature of \gampen{} is that it uses a Spatial Transformer Network (STN; \citealp{NIPS2015_33ceb07b}) module upstream of CNN, which crops each input image to an optimal size so that the performance of the downstream CNN module is maximized.
Both its STN and CNN modules can be trained jointly without any additional supervision, and this greatly reduces the work of choosing the best cutout size for each galaxy image. 

In training, validating, and testing our neural networks, we use high-quality imaging data available from the Hyper Suprime-Cam (HSC) Wide survey public data release 3 \citep{2021arXiv210813045A}, which has a sky coverage of $\sim 1400$ deg$^2$ in five broad-band filters ($grizy$) with an average depth of $\sim 26$, an average seeing of $\sim 0.7$ arcsec and a pixel scale of $0.168$ arcsec per pixel.
We first train our models on a large amount of simulated data created by GalSim \citep{2015A&C....10..121R} that has comparable parameter ranges as real HSC data, and then we use transfer learning \citep[e.g.][]{2018MNRAS.479..415A, 2019MNRAS.484...93D, 2020ApJ...895..112G, 9709840} to fine-tune the parameters of our models on a small amount of real HSC data.
Although we chose to optimize our models to HSC, one can easily use transfer learning again with our trained models for other imaging data.

We organize this paper as follows: 
In Section~\ref{sec:data}, we present the real HSC data and the simulated data that we used in training, validating and testing our neural network models.
In Section~\ref{sec:ml_frame}, we briefly introduce the two main components of our model---PSFGAN and \gampen{}---followed by an overview of our approach that we use to quantitatively study the morphological parameters of the AGN host galaxy.
The details of data preparation and network training are discussed in Section~\ref{sec:training}.
Then we comprehensively analyze our model performance in Section~\ref{sec:perfo}.
In Section~\ref{sec:summary}, we summarize our results and main conclusions and leave comments on the planned future works. 
Throughout this paper, we use a $\Lambda$ CDM cosmology model of $H_0 = 70 \ {km} \ s^{-1} \ {Mpc}^{-1}$, $\Omega_m = 0.3$, and $\Omega_{\Lambda} = 0.7$.

\section{Data} \label{sec:data}
\subsection{Hyper Suprime-Cam Galaxies} \label{sec:hsc_gal}
We use high-quality imaging data from the Hyper Suprime-Cam (HSC) Subaru Strategic Program Public Data Release 3 \citep{2019PASJ...71..114A}, in its five wide field bands: \gb{}, \rb{}, \ib{}, \zb{}, and \yb{}.
Additionally, we divide our dataset into five bins based on the photometric redshift of the sources: \gbin{} ($0 < z < 0.25$), \rbin{} ($0.25 < z < 0.5$), \ibin{} ($0.5 < z < 0.9$), \zbin{} ($0.9 < z < 1.1$) and \ybin{} ($1.1 < z < 1.4$).
In each redshift bin, we use images from only one band---namely, \gb{} for $0 < z < 0.25$, \rb{} for $0.25 < z < 0.5$, \ib{} for $0.5 < z < 0.9$, \zb{} for $0.9 < z < 1.1$ and \yb{} for $1.1 < z < 1.4$.
This ensures that we are targeting approximately the same rest-frame wavelengths in each redshift bin.
Table \ref{tb:band_vs_zbin} shows this correspondence between redshift bins and bands for rest-frame emission at $\lambda_{rest} = 450$ nm.
We optimize a unique ML model for galaxies in each redshift bin, in order to improve the overall model performance.
Galaxies beyond $z=1.4$ are not considered in this paper, because the apparent size of host galaxies at $z>1.4$ become comparable to the PSF size, and the signal-to-noise ratios are too low.

\begin{deluxetable}{cccc}[htbp]
\tablecaption{Correspondence between Redshift Bins and Bands for \\ $\lambda_{rest} = 450$ nm \label{tb:band_vs_zbin}}
\tablecolumns{4}
\tablehead{
\colhead{Band} & \colhead{Coverage (nm)\tablenotemark{a}} & \colhead{Redshift Bin} &  \colhead{$\lambda_{obs}$ (nm)\tablenotemark{b}}   
}

\startdata
    \hline
    \hline
    \gb & $400 - 550$ & \gBin{} ($0 < z < 0.25$)  & $450 - 562.5$ \\
    \rb & $550 - 695$ & \rBin{} ($0.25 < z < 0.5$)  & $562.5 - 675$ \\
    \ib & $695 - 845$ & \iBin{} ($0.5 < z < 0.9$)  & $675 - 855$ \\
    \zb & $845 - 930$ & \zBin{} ($0.9 < z < 1.1$)  & $855 - 945$ \\
    \yb & $930 - 1070$ & \yBin{} ($1.1 < z < 1.4$)  & $945 - 1080$ \\
\enddata
\tablenotetext{a}{Wavelengths covered by this HSC band.}
\tablenotetext{b}{
The range of observed wavelengths corresponding to $\lambda_{rest} = 450$ nm across this redshift bin.
}
\end{deluxetable}

To select appropriate galaxies for our study, we used the PDR 3 catalog produced using forced photometry on coadded images, filtered according to a few characteristic and quality flags.
Specifically, we used the \texttt{extendedness\_value} flag to select extended sources and we constrain our sample to $m<23$~mag in each corresponding band so that the contamination from stars is minimal\footnote{As described in \href{https://hsc-release.mtk.nao.ac.jp/doc/index.php/star-galaxy-separation__pdr3/}{https://hsc-release.mtk.nao.ac.jp/doc/index.php/star-galaxy-separation\_pdr3/}.} and \texttt{extendedness\_value} can therefore be used as a reliable indicator to separate galaxies and stars.
Throughout this paper, we use photometric redshifts calculated using Mizuki \citep{mizuki}, a fitting algorithm with Bayesian priors, discarding all sources with \texttt{photoz\_risk\_best} parameter is greater than $0.1$. (\texttt{photoz\_risk\_best} indicates the risk of the photo-z being outside of the range $z\_true \pm 0.15(1+z\_true)$, which ranges from $0$ (extremely safe) to $1$ (extremely risky)).
Furthermore, we select only primary sources (\texttt{isprimary} is \texttt{TRUE}, meaning they are not blended with other sources and are in the inner region of a coadd patch and tract).
We also use \texttt{pixelflags}, \texttt{is\_clean\_centerpixels} and \\ \texttt{is\_clean\_allpixels} in each corresponding band to filter out images with unusable, contaminated, interpolated, suspect, saturated, or bad pixels, or those that are not clean at the center of the image and everywhere.
Finally, we apply a constraint of \texttt{m\_\{filter\}\_blendedness\_abs} less than 0.01 for each source in the corresponding band, so that mergers and blending galaxies are excluded from our sample.

\begin{figure}
\figurenum{1}
\gridline{\fig{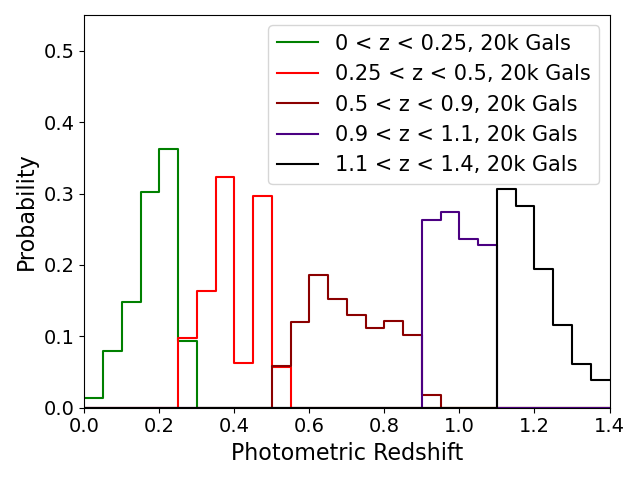}{0.48\textwidth}{}}
\vspace*{-0.6cm}
\gridline{\fig{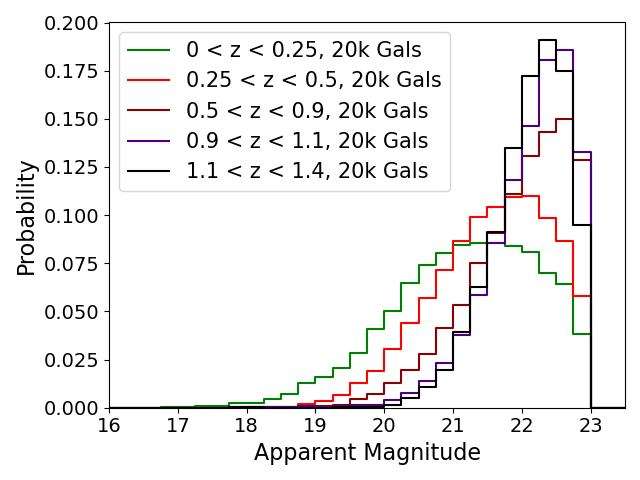}{0.48\textwidth}{}}
\vspace*{-0.6cm}
\gridline{\fig{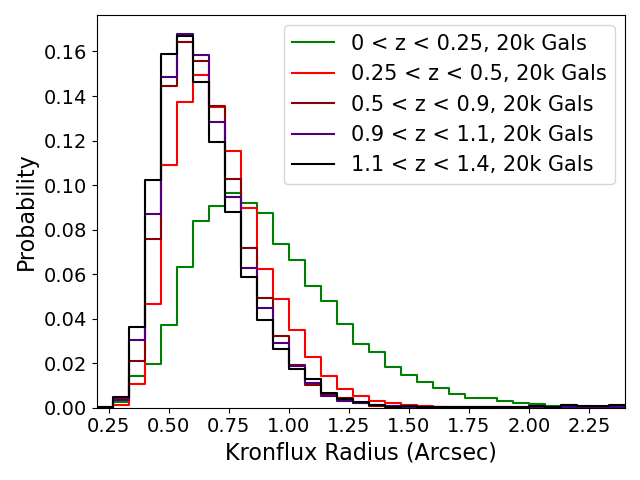}{0.48\textwidth}{}}
\vspace*{-0.6cm}
\caption{Photometric redshift (\textit{top}), magnitude in corresponding band (\textit{middle}) and Kron radius (\textit{bottom}) distributions of HSC galaxies used to train, validate and test our ML models: 20,000 galaxies in each redshift bin. Each histogram is normalized separately to unity.
\label{fg:hsc_gal_pre}}
\end{figure}

Once all constraints have been applied, in each redshift bin we randomly select 20,000 galaxies for training, validating, and testing our ML models.
The photometric redshift, apparent magnitude and Kron radius\footnote{Defined as the $sqrt (major\_axis \times minor\_axis)$.} distributions of our selected data are shown in Figure \ref{fg:hsc_gal_pre}.
For each galaxy, we use the \href{https://hsc-release.mtk.nao.ac.jp/das_cutout/pdr3/}{Image Cutout Tool} provided by HSC to prepare 185 pixel (about $31\arcsec$ in HSC Wide) square cutouts in the corresponding band.
Note that we use the same cutout size for the training and testing of PSFGAN. 
When we invoke \gampen{}, galaxy images will be cropped to various smaller sizes depending on redshift bin, with the largest size being $30\arcsec$---that is why we choose $31\arcsec$ as the downloading size (see Section~\ref{sec:init_training} and Section~\ref{sec:trans_learning} for details).
In over $99.9\%$ of cases (also reflected in Figure \ref{fg:hsc_gal_pre}), the chosen cut-out size is larger than $10$ times the size (Kron radius) of the galaxy.
These downloaded galaxy images are first sent to GALFIT for fitting (Appendix \ref{sec:ap:galfitting}), so that we can create truth labels for training and testing \gampen{}.

\subsection{Hyper Suprime-Cam Stars} \label{sec:hsc_star}
To create artificial AGN point sources (Section~\ref{sec:prep_trans}), we first select a few bright stars, using the following flags from the PDR 3 catalog: 
\begin{enumerate}
    \itemsep0em 
    \item \texttt{cmodel\_mag} is less than 23;
    \item \texttt{extendedness\_value} is 0 in all five bands;
    \item \texttt{photoz\_best} is less than 0.1;
    \item \texttt{photoz\_risk\_best} is less than 0.2;
    \item \texttt{isprimary} is \texttt{TRUE};
    \item Each of \texttt{sdsscentroid\_flag}, \texttt{cmodel\_flag} and \texttt{pixelflags} is \texttt{FALSE};
    \item Each of \texttt{is\_clean\_centerpixels} and \texttt{is\_clean\_allpixels} is \texttt{TRUE}.
\end{enumerate}
After applying these constraints, for each star image, we sum the flux within a 20-pixel square region at the image center.
We then calculate the fraction between this central-region flux to the total flux of the image and discard any star image with this fraction less than 0.5.
This step effectively filters out images that have other sources of comparable brightness relative to the central star.
In the resulting star images, we randomly select 360 images in the corresponding band for each redshift bin.
We again used a square size of 185 pixels ($\sim 31\arcsec$) when preparing these star images. 

\subsection{Simulated Galaxies} \label{sec:sim_gal}

\begin{deluxetable*}{ccccccc}[htbp]
\tablecaption{Parameter Ranges for Simulated Galaxies in Each Redshift Bin\label{tb:sim_gal}}
\tablecolumns{7}
\tablehead{
\colhead{Component} & \colhead{\sersic\ Index} & \colhead{Half-Light Radius\tablenotemark{a}} & \colhead{Flux }& \colhead{Magnitude\tablenotemark{b}} & \colhead{Axis Ratio} & \colhead{Position Angle} \\ 
\colhead{} & \colhead{} & \colhead{(Arcseconds)} &  \colhead{(ADUs)} & \colhead{} & \colhead{} & \colhead{(Degrees)}
}
\startdata
    \hline
    \hline
    \multicolumn{7}{c}{37,500 Single-Component Galaxies\tablenotemark{c}, images in the corresponding band\tablenotemark{d}} \\
    \hline
    && 0.1--5.0
    &30--$1.35\times 10^{5}$& 14.1--23.3 &&\\
    && 0.1--3.0
    &30--$3\times 10^{4}$& 15.8--23.3 &&\\
    Single & 0.8--1.2 or 3.5--5.0 & 0.1--2.0 & 30--$2\times 10^{4}$  & 16.2--23.3 & 0.25--1.0 & $-90.0$--$90.0$ \\
    && 0.1--2.0 
    &30--$1\times 10^{4}$ & 17.0--23.3 &&\\
    && 0.1--2.0 
    &30--$8\times 10^{3}$ & 17.2--23.3 &&\\
    \hline
    \hline
    \multicolumn{7}{c}{112,500 Double-Component Galaxies, images in the corresponding band} \\
    \hline
    Disk & 0.8--1.2 & same as ``Single''\tablenotemark{e}
    & 0--1\tablenotemark{f} && 0.25--1.0 & $-90.0$--$90.0$ \\
    && 0.1--3.0 
    &&&&\\
    && 0.1--2.0 
    &&&&\\
    Bulge & 3.5--5.0 & 0.1--1.5 
    & 1 $-$ Disk Frac.  && 0.25--1.0 & Disk Comp.$\pm$\,(0,15)\tablenotemark{g}\\
    && 0.1--1.5 
    &&&&\\
    && 0.1--1.5
    &&&&\\
    \hline
\enddata
\tablenotetext{a}{For either the single or the double component galaxies, we display parameter ranges used by each of the five redshift bins: \gbin{}, \rbin{}, \ibin{}, \zbin{}, and \ybin{}, from top to bottom. The same format is used for the flux and magnitude columns as well.}
\tablenotetext{b}{Calculated based on HSC photometric zeropoint: $Magnitude = -2.512 \times log(Flux_{ADUs}) + 27.0$. Double component galaxies have the same magnitude range as for single component galaxies in each redshift bin.}
\tablenotetext{c}{Divided equally into disks and bulges.}
\tablenotetext{d}{Band \gb{}, \rb{}, \ib{}, \zb{} or \yb{} for the \gbin{} ($0<z<0.25$), \rbin{} ($0.25<z<0.5$), \ibin{} ($0.5<z<0.9$), \zbin{} ($0.9<z<1.1$), or \ybin{} ($1.1<z<1.4$) redshift bin, respectively.}
\tablenotetext{e}{The range of half-light radius for the disk component in double-component galaxies is set equal to the range of the same parameter for single component galaxies, in each redshift bin respectively.}
\tablenotetext{f}{Only the fractional flux is mentioned here. The flux for the disk (bulge) component equals to the total flux multiplied by the disk (bulge) fractional flux. The total flux is chosen from the same range as for single component galaxies, in each redshift bin respectively.}
\tablenotetext{g}{The bulge position angle differs from the disk position angle by a randomly chosen value between $-15^\circ$ and $+15^\circ$.}
\tablecomments{This table shows details of parameter ranges of simulated galaxies in each of the five redshift bins. All parameters are drawn from uniform distributions except the bulge position angle in double-component galaxies. We choose parameter ranges such that, given the size of our simulated data, they resemble real galaxies in the corresponding redshifted bands. See Section~\ref{sec:sim_gal} for more details.}
\end{deluxetable*}

For the initial training of our PSFGAN and \gampen{} models, we used GalSIM to create $150,000$ simulated galaxies in the corresponding band for each redshift bin.
We chose simulation parameters (e.g., half-light radius and flux) that match observed ranges in most real galaxies in each redshift bin, according to \citet{binney_and_merrifield}.
Specifically, $75\%$ of simulated galaxies ($112,500$) have both a disk and a bulge.
To make a set of various light profiles, we used a \sersic{} index $n$ between $0.8$ and $1.2$ for the disk component and a \sersic{} index $n$ between $3.5$ and $5.0$ for the bulge component\footnote{All these parameters are drawn from uniform distributions---see the note in Table \ref{tb:sim_gal} for details.}, in contrast to using fixed \sersic{} indices ($n=1$ for disk and $n=4$ for bulge).
The remaining $25\%$ of simulated galaxies ($37,500$) have a single component, which is either a disk or a bulge (created with the same \sersic{} index ranges as for double-component galaxies, $18,750$ for each type).
For double component galaxies, we chose an axis ratio $b/a$ between $0.25$ and $1.0$ independently for its disk and bulge components (see, for instance, \citealp{1997AJ....114.2219O, Lee_2006}).
In addition, we randomly chose a position angle between $-90^\circ$ and $90^\circ$ for the disk component, then chose a random bulge orientation within $\pm 15^\circ$ of the disk orientation.
We also drew the bulge-to-total light ratio $L_{B}/L_{T}$ uniformly between $0$ and $1$.
Single-component galaxies were created using the same ranges of parameters used for the disk component in double-component galaxies (except for flux and magnitude, which use the same ranges for the entire double component galaxies; see Table \ref{tb:sim_gal}).
Details of all parameter ranges are presented in Table \ref{tb:sim_gal}.
Note that our chosen ranges for the half-light radius and the flux/magnitude are different across redshift bins, and these ranges are, in general, more expansive\footnote{That is, parameter ranges for our simulated galaxies are wider in comparison to parameter ranges for real HSC galaxies we use.} than what was observed in our real HSC data (Figure \ref{fg:hsc_gal_pre}).
Finally, we use the same square cutout size of 185 pixels ($\sim 31\arcsec$) for each of the simulated galaxies.

To closely resemble real HSC galaxies, as in our previous works \citep{2022ApJ...935..138G, Ghosh_2023, 2023ApJ...944..124T}, we convolve each simulated galaxy with a HSC instrumental PSF.
For each redshift bin, we first select 50 PSFs in the corresponding band from random locations in HSC Wide PDR3, using the \href{https://hsc-release.mtk.nao.ac.jp/psf/pdr3/}{HSC PSF Picker Tool}.
Each simulated galaxy is convolved with a randomly selected PSF from this chosen group of 50 PSFs.
At the end, we add noise using \href{https://galsim-developers.github.io/GalSim/\_build/html/noise.html}{GalSim's inherent noise module}\footnote{Parameters we used to resemble real HSC data: \texttt{type}: CCD; \texttt{gain}: 3.0; \texttt{read\_noise}: 4.5.} after PSF convolution.

\section{ML Frameworks and Approach} \label{sec:ml_frame}
\subsection{Introduction to PSFGAN} \label{sec:intro_psfgan}

\begin{figure*}
\figurenum{2}
\gridline{\fig{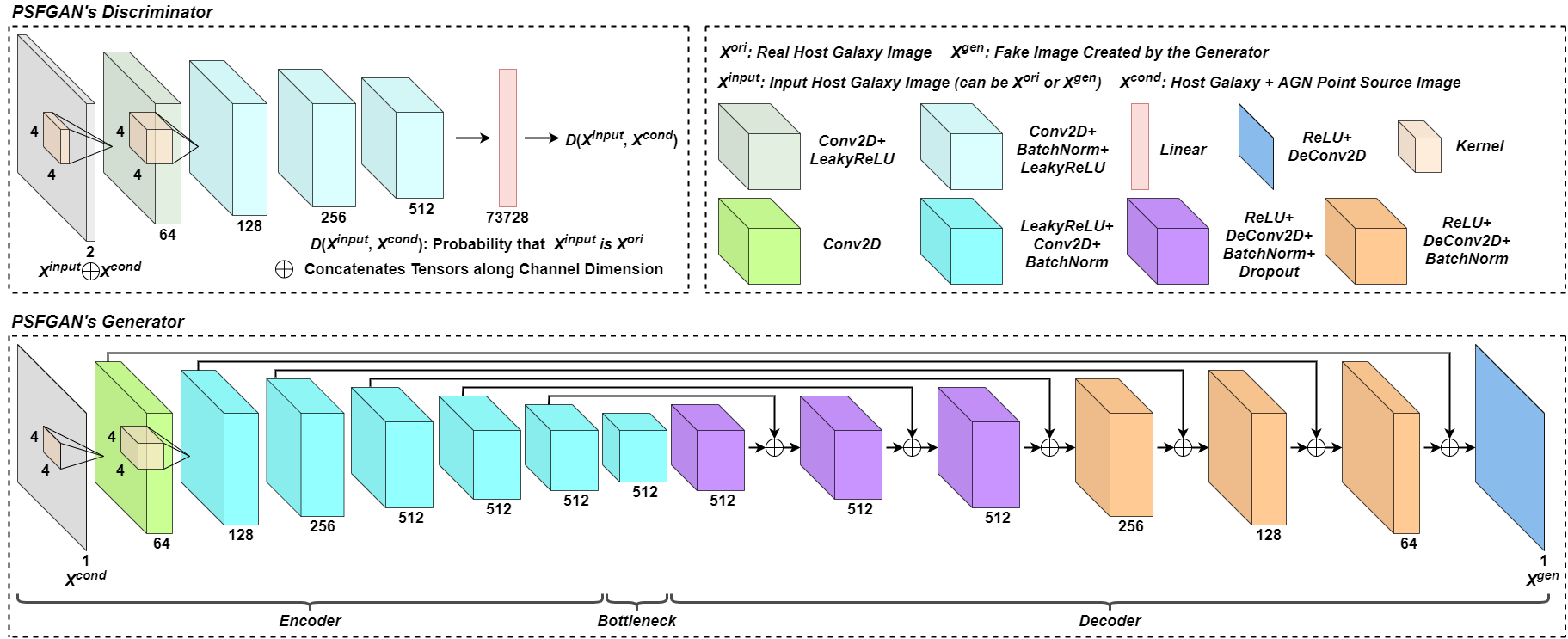}{1.0\textwidth}{}}
\vspace*{-0.6cm}
\caption{Architecture of PSFGAN used in our experiment. 
PSFGAN consists of two neural networks: the generator and the discriminator.
The generator takes an image of the host galaxy + AGN point source and tries to create a fake image of the host galaxy alone, while the discriminator takes an unknown input image of galaxy and tries to determine whether it's a real image or a fake one created by the generator.
Both networks use convolutional layers to compress the input image into a hidden representation of smaller dimensions, and the generator also uses de-convolutional layers to create its output based on the hidden representation --- we thus divide the generator into three parts and refer to them as the encoder, the bottleneck, and the decoder.
The number of channels (the ``thickness'' of convolutional and de-convolutional layers) or neurons (for fully-connected layers) is denoted below each layer (both are not to scale).
All kernels use the same size and this number is denoted next to the corresponding yellow box; only the first two kernels are shown for each of the two networks. 
Conv2D, DeConv2D, ReLU, LeakyReLU, and BatchNorm refer to Convolutional Layers, De-Convolutional Layers, Rectified Linear Units, Leaky Rectified Linear Units, and Batch Normalization Layers, respectively.
See Section~\ref{sec:init_training} for more details (including PSFGAN's loss function).
\label{fg:psfgan_arch}}
\end{figure*}

We now introduce the first ML framework that we use in this paper---PSFGAN \citep{2018MNRAS.477.2513S}---a GAN \citep{goodfellow2014generative} designed to remove bright AGN point sources.
Briefly, PSFGAN consists of two neural networks.
The first network (``generator'') takes an image of host galaxy + AGN point source (that is, an active galaxy --- can be a simulated one or real one) and makes a reasonable guess of what the host galaxy alone should look like when generating its output.
The second network (``discriminator'') tries to determine whether an unknown input image is a real (original) image or a fake (recovered) image created by the generator.
Both the generator and the discriminator are trained simultaneously, and the performance of both networks can be improved at the same time.
At the end of the training, the discriminator can find tiny 
differences between the recovered and original images.
In turn, the generator must be able to generate fake images of the host galaxy that closely resemble realistic ones.
Interested readers are referred to \citealp{2018MNRAS.477.2513S} for an comprehensive overview of PSFGAN technical details and to \citealp{goodfellow2014generative} for an excellent introduction to GAN in general\footnote{Strictly speaking, PSFGAN is a Conditional GAN \citep{mirza2014conditionalgenerativeadversarialnets}; that is, its generator generates fake images using some extra information (i.e., images of host galaxy + AGN point source), unlike the ``vanilla GAN'' originally described in \citealp{goodfellow2014generative}, which generates images from noise.}.

Figure \ref{fg:psfgan_arch} shows the architecture of the PSFGAN we constructed, which is identical to that used by \citet{2018MNRAS.477.2513S} except that we removed the last layer from generator's encoder and the first layer from generator's decoder\footnote{That is, with respect to the original PSFGAN's generator, we took out the innermost two layers adjacent to the ``bottleneck'' --- see Figure \ref{fg:psfgan_arch}.}.
Since the original PSFGAN \citep{2018MNRAS.477.2513S} took 424 pixel square inputs while our PSFGAN takes 185 pixel square inputs instead, this modification ensures that the ``bottleneck'' of our generator remains about the same size of the original PSFGAN (i.e., the hidden representation has about the same dimensions as in the original PSFGAN; see Figure \ref{fg:psfgan_arch} caption).
The discriminator remains unchanged since it always maps an input of (input galaxy, host galaxy + AGN point source) to an output value (a scalar) that represents the probability that the input galaxy is a real one according to its belief.

\subsection{Introduction to \gampen{}} \label{sec:intro_gampen}
The second ML framework we use in this paper is the Galaxy Morphology Posterior Estimation Network (\gampen{}), which has two modules: Spatial Transformer Network (STN) module and Convolutional Neural Network (CNN) module.
We briefly introduce each of them in this subsection.
For a comprehensive discussion of these modules, the reader is referred to the original \gampen{} paper \citep{2022ApJ...935..138G}.

\begin{figure*}
\figurenum{3}
\gridline{\fig{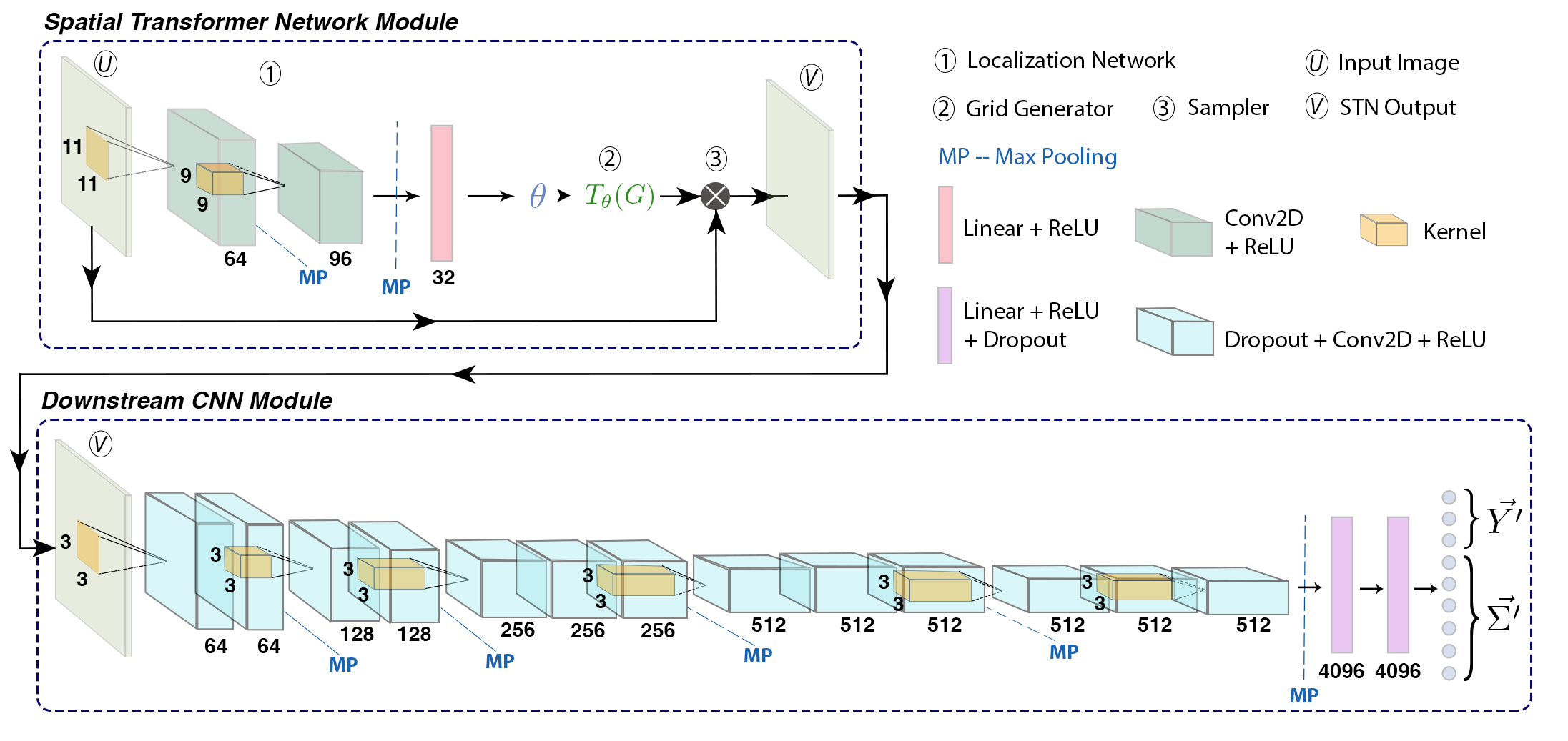}{1.0\textwidth}{}}
\vspace*{-0.6cm}
\caption{Architecture of \gampen{}, from Figure 3 in \citealp{2022ApJ...935..138G}.
The Galaxy Morphology Posterior Estimation Network (\gampen{}) consists of two modules: an upstream STN module, followed by a downstream CNN module.
The CNN module estimates posterior distributions of morphological parameters of the input galaxy, while the STN module learns to perform a just right affine transformation (we constrain this to be cropping only; see Section~\ref{sec:training}) of the input galaxy such that the CNN module performance is maximized.
The number of filters (or neurons) is denoted below each layer.
The kernel size used by filters in each layer is denoted next to the corresponding yellow box; all filters in each layer use the same kernel size.
Only one kernel is shown per set of convolutional layers; all other layers in the set have kernels of the same size.
Conv2D and ReLU refer to Convolutional Layers and Rectified Linear Units, respectively.
See \citealp{2022ApJ...935..138G} for more details.
\label{fg:gampen_arch}}
\end{figure*}

\subsubsection{Introduction to \gampen{} STN Module} \label{sec:intro_stn}
As shown in Figure \ref{fg:gampen_arch}, \gampen{} consists of an upstream STN module and a downstream CNN module.
The STN module performs an affine transformation\footnote{Translation, reflection, rotation, scaling (cropping) or shear mapping.} (in our case, we constrain this affine transformation to be cropping-only) on each input image, and then it passes the transformed image to the CNN module for estimating the posterior distributions.
During training, the STN module learns the best way to make this transformation so that the performance of the CNN module is optimized.
Both STN and CNN are trained together, and we use back-propagation to pass errors back to the STN, therefore no additional supervision is needed for the STN to learn how to make the best transformation.

The STN module was introduced when \gampen{} was designed to make the framework applicable to galaxies spanning a wide range of redshift and brightness. In a scenario where input galaxies vary greatly in size, one must find a way to choose an appropriate cutout size for each galaxy in a way such that the entire galaxy-to-be-studied is contained in the cutout and the galaxy occupies a reasonable fraction of the image (in other words, there are not too many secondary objects).
Choices of these cutout sizes are traditionally linked to labeled galaxy sizes, which, in the case we want to use our models to estimate galaxy sizes, are usually not available.
Therefore, the STN was introduced to automatically crop each input galaxy to a proper size, without the need for any labeled galaxy sizes.

\subsubsection{Introduction to \gampen{} CNN Module} \label{sec:intro_cnn}
The downstream CNN module estimates posterior distributions of the bulge-to-total light ratio ($L_B/L_T$), the half-light radius ($R_e$) and the total flux ($F$)\footnote{\gampen{} can be used to estimate the posterior distribution for an arbitrary number of parameters.} of input galaxies.
In doing so, there are two primary sources of error: the aleatoric uncertainty, which refers to the intrinsic errors introduced by input images (blurring, noise, instrumental limitation, etc., which can't be reduced given the data), and the epistemic uncertainty, which is associated with the errors introduced by the model itself (the complexity of our models, the lack of training data, etc., which can be reduced with optimal training of the model).
Here we show how \gampen{} addresses both uncertainties.

For each time an input image is passed, \gampen{} predicts a multivariate Gaussian distribution $\mathcal{N}(\boldsymbol{\mu}, \boldsymbol{\Sigma})$, where $\boldsymbol{\mu}$ is a three-dimensional vector containing the predicted mean of $L_B/L_T$, $R_e$ and $F$, and $\boldsymbol{\Sigma}$ is the associated 3 by 3 covariance matrix.
Since we only have access to the true values (not including any covariance information) from our training labels, we train \gampen{} to minimize the negative log-likelihood of the output parameters and learn $\boldsymbol{\Sigma}$ from training.
This covariance matrix represents the aleatoric uncertainty.

After training \gampen{}, we pass each input image multiple times into the trained framework for inference.
We use the Monte-Carlo Dropout technique \citep{Srivastava2014Dropout:Overfitting} in this process: random neurons from the network are dropped\footnote{Each neuron is dropped with an independent probability, $p$, known as the dropout rate.} right before each time the input is passed.
Therefore, we are passing the input to a slightly different network every time.
For a certain input image $\boldsymbol{\hat{X}}_n$, we pass it through the network $T$ times.
For the $t^{th}$ forward pass, we draw a sample $\boldsymbol{\hat{Y}_{n,t}}$ (a 3-dimensional vector containing a sampled value for each of the three parameters) from the predicted multivariate normal distribution $\mathcal{N}\left(\hat{\boldsymbol{\mu}}_{n,t},\boldsymbol{\hat{\Sigma}}_{n,t}\right)$.
We then collect the $\boldsymbol{\hat{Y}_{n,t}}$ samples for all $t$, and use them to generate the predicted posterior distribution of $\boldsymbol{\hat{Y}_{n}}$ that corresponds to the input image $\boldsymbol{\hat{X}}_n$.
Note that epistemic uncertainties are captured when we draw the predicted posterior distribution of $\boldsymbol{\hat{Y}_{n}}$ (from multiple forward passes), while aleatoric uncertainties are captured by each covariance matrix $\boldsymbol{\hat{\Sigma}_{n,t}}$ (from each single forward pass). 
In this way, we incorporate both uncertainties in our \gampen{} predictions.

\subsection{Overview of Our Approach} \label{sec:overv_appro}

\begin{figure*}
\figurenum{4}
\gridline{\fig{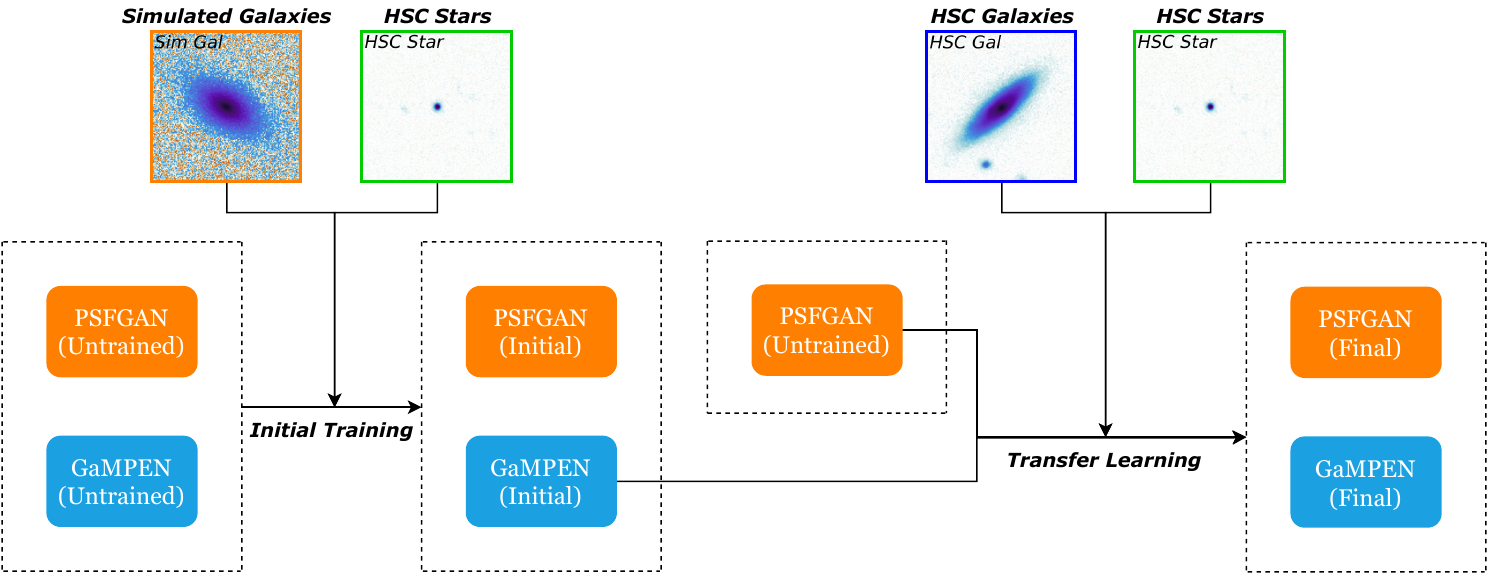}{1.0\textwidth}{}}
\vspace*{-0.6cm}
\caption{Diagram showing overview of PSFGAN + \gampen{} training process.
There are two separate phases: initial training and transfer learning.
For the initial training phase, we used simulated galaxies and real HSC stars with an untrained PSFGAN and an untrained \gampen{}.
During the transfer learning phase, we take another untrained PSFGAN and the previously trained \gampen{}, along with real HSC galaxies and stars.
Our detailed approach in each phase is similar; see Figure \ref{fg:appro_details} for details.
\label{fg:appro_overall}}
\end{figure*}

\begin{figure*}
\figurenum{5}
\gridline{\fig{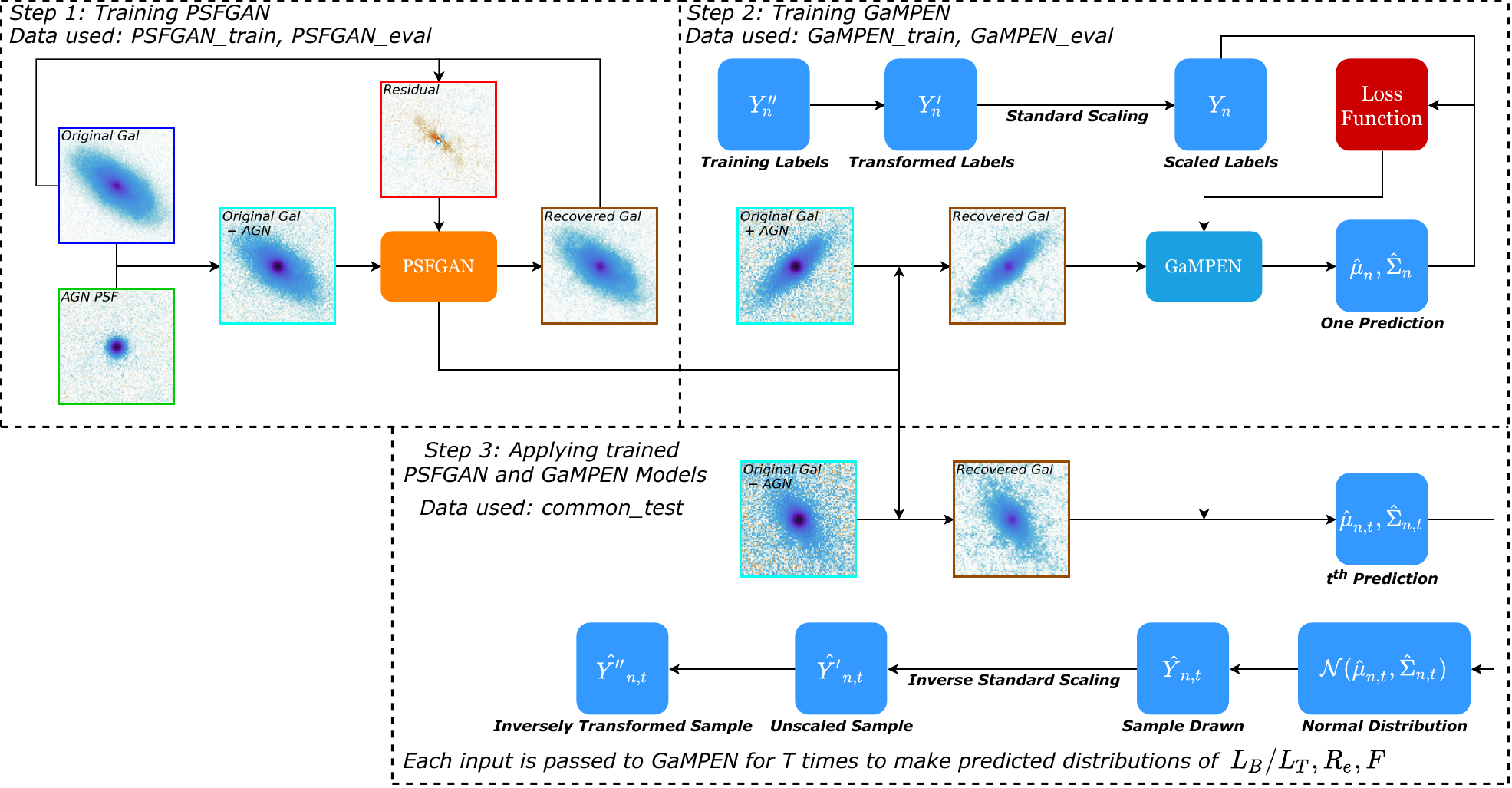}{1.0\textwidth}{}}
\vspace*{-0.6cm}
\caption{Diagram showing details of our training process.
In both the initial training and the transfer learning phases, we use the same 3-step approach summarized here, with simulated data in the initial training phase and real HSC data in the transfer learning phase, respectively.
{\it Step 1:} We use simulated (or real HSC) galaxies with known morphological parameters from the simulation (or from GALFIT fitting) in the training and validation sets of PSFGAN.
Artificial AGN point sources are added, and both the original galaxy and the original + AGN are fed to PSFGAN for training.
In the same way, we create AGN point sources and original + AGN images for galaxies in the training and validation sets of \gampen{} and the common test set.
In the training process, PSFGAN compares each recovered galaxy (i.e., its own guess) to the original galaxy and tries to adjust itself so that the difference between these two is minimal.
It does this by incorporating a residual term into the loss function of its generator.
Once PSFGAN is trained, we apply it on the training and validation sets of \gampen{} and the common test set. 
{\it Step 2:} We use recovered galaxies in the training and validation sets of \gampen{} (from the PSFGAN application at the end of the previous step).
Training labels associated to each galaxy are transformed and then scaled so that all three parameters have similar numerical ranges.
For each input galaxy, \gampen{} compares its output with the associated scaled labels (by minimizing their negative log-likelihood) during training. 
{\it Step 3:} Each recovered galaxy in the common test set is passed to the trained \gampen{} for $T$ times.
During each feed-forward pass, we draw a sample $\boldsymbol{\hat{Y}_{n,t}}$ from the predicted multivariate normal distribution $\mathcal{N}\left(\hat{\boldsymbol{\mu}}_{n,t},\boldsymbol{\hat{\Sigma}}_{n,t}\right)$.
The sample is then unscaled and inversely transformed to $\boldsymbol{\hat{Y''}_{n,t}}$.
At the end, we aggregate all $\boldsymbol{\hat{Y''}_{n,t}}$'s to make the predicted posterior distributions for $L_B/L_T$, $R_e$ and $F$.
See Section~\ref{sec:overv_appro} for a more comprehensive discussion.
\label{fg:appro_details}}
\end{figure*}

In this subsection, we give an overview of our approach to build a composite framework to estimate posterior distributions of active galaxy morphological parameters in the HSC Wide survey. 
The procedures discussed below are for a generic redshift bin, since we use the same approach in each bin.

Generally speaking, our approach can be divided into two phases: initial training, and transfer learning.
During the initial training phase, we use simulated galaxies (Section~\ref{sec:sim_gal}) and real HSC stars (Section~\ref{sec:hsc_star}) to train a PSFGAN and a \gampen{} from scratch.
Once this is done, we start the transfer learning phase, in which we use real HSC galaxies (Section~\ref{sec:hsc_gal}) and the same set of stars to fine-tune the previously trained \gampen{}.
Figure \ref{fg:appro_overall} sketches both phases.
Note that in the transfer learning phase, we still train a PSFGAN from scratch (that is, we only apply the transfer learning technique to \gampen{} models).
The advantage of this combined approach is that we can generate as many simulated galaxies as we want (only limited by our computing resources)---which makes it possible to use a smaller set of real galaxies in the transfer learning phase.
Should we only have one phase, we must be able to provide a much larger training set of real HSC galaxies with known morphological parameters---which is usually the bottleneck in similar ML approaches, no matter whether one uses labeled catalogs from the literature or fits light profiles to create labels by its own.

Within each phase, the detailed procedures are very similar, so we depict the procedures for a generic phase in Figure \ref{fg:appro_details}.
We can briefly split each phase into three steps: Step 1 (PSFGAN training), Step 2 (\gampen{} training), and Step 3 (application of trained PSFGAN and \gampen{} models).
Besides this, we split all galaxies (regardless of whether we are using simulated or real HSC galaxies) into five subsets: the training set of PSFGAN, the validation set of PSFGAN, the training set of \gampen{}, the validation set of \gampen{}, and the common test set.
The first two subsets are used in Step 1, the third and fourth subsets are for Step 2, and the last set is for Step 3.

In Step 1, we use real HSC stars to create artificial AGN point sources, and then add them to original galaxies to create original + AGN images, for galaxies in all five subsets mentioned above (this AGN creation procedure is discussed in detail in Section~\ref{sec:prep_trans}).
To begin training, we feed the original galaxy and original + AGN image to an untrained PSFGAN. The network makes its own guess of the recovered host galaxy, which is then compared to the original galaxy.
The calculated residual is sent back to the network, which learns from it to make a better recovered host galaxy (that is, closer to the original one) next time.
In this step, we used galaxies from the training and validation sets of PSFGAN.
Once we ensure that PSFGAN is properly trained, we apply it on the original + AGN images in the remaining three subsets: training and validation sets of \gampen{} and the common test set.

We then perform Step 2, during which we train the \gampen{} models.
Before starting the training process, the training labels (either from simulation parameters or GALFIT fitting) associated with each galaxy are first transformed and then scaled in such a way that all three parameters have similar numerical ranges (discussed in Section~\ref{sec:prep_trans} as well).
This step ensures that none of the three parameters can make a disproportionate contribution to the loss function.
On the other hand, we take the \gampen{} model---either an untrained one (for the initial training phase), or a previously trained one (for the transfer learning phase)---and feed the PSFGAN-recovered galaxy to it.
\gampen{} makes its own prediction, which is compared to the scaled training labels by calculating the loss function.
The loss is sent back to \gampen{} so it can learn from its own mistake.
We used galaxies from the training and validation sets of \gampen{} in this step.

Finally, in Step 3 
we use recovered galaxies in the common test set, which were made available to us via the application of PSFGAN at the end of Step 1.
We pass each recovered galaxy to the trained \gampen{} for $T$ times.
During each feed-forward pass, \gampen{} makes a unique prediction, which corresponds to a unique multivariate normal distribution, and from that we draw a sample (please also refer to the end of Section~\ref{sec:intro_cnn}). 
Each sample drawn is unscaled and inversely transformed.
After that, we aggregate all the samples (from all $T$ feed-forward passes) to make the predicted posterior distributions of the three morphological parameters.

It is worth mentioning that as described above, we used a linked approach---meaning that we train \gampen{} models on PSFGAN-recovered images and test it on PSFGAN-recovered images as well.
In the past, including our previous work \citep{2023ApJ...944..124T}, we tried to use a mutually independent approach---i.e., we trained \gampen{} models on original images and tested them on PSFGAN-recovered images.
We empirically tested that this mutually independent approach results in a much lower model performance than the linked one.
The underlying reason is that, relative to original images, PSFGAN can introduce faint but unnatural patterns at the center of recovered images.
When making predictions, these patterns can confuse a \gampen{} model that is trained on original images only.
Therefore, we adopted the linked approach in this paper.

\section{Model Training} \label{sec:training}

\subsection{Data Prepared and Transformed} \label{sec:prep_trans}

\begin{figure}
\figurenum{6}
\epsscale{1.2}
\plotone{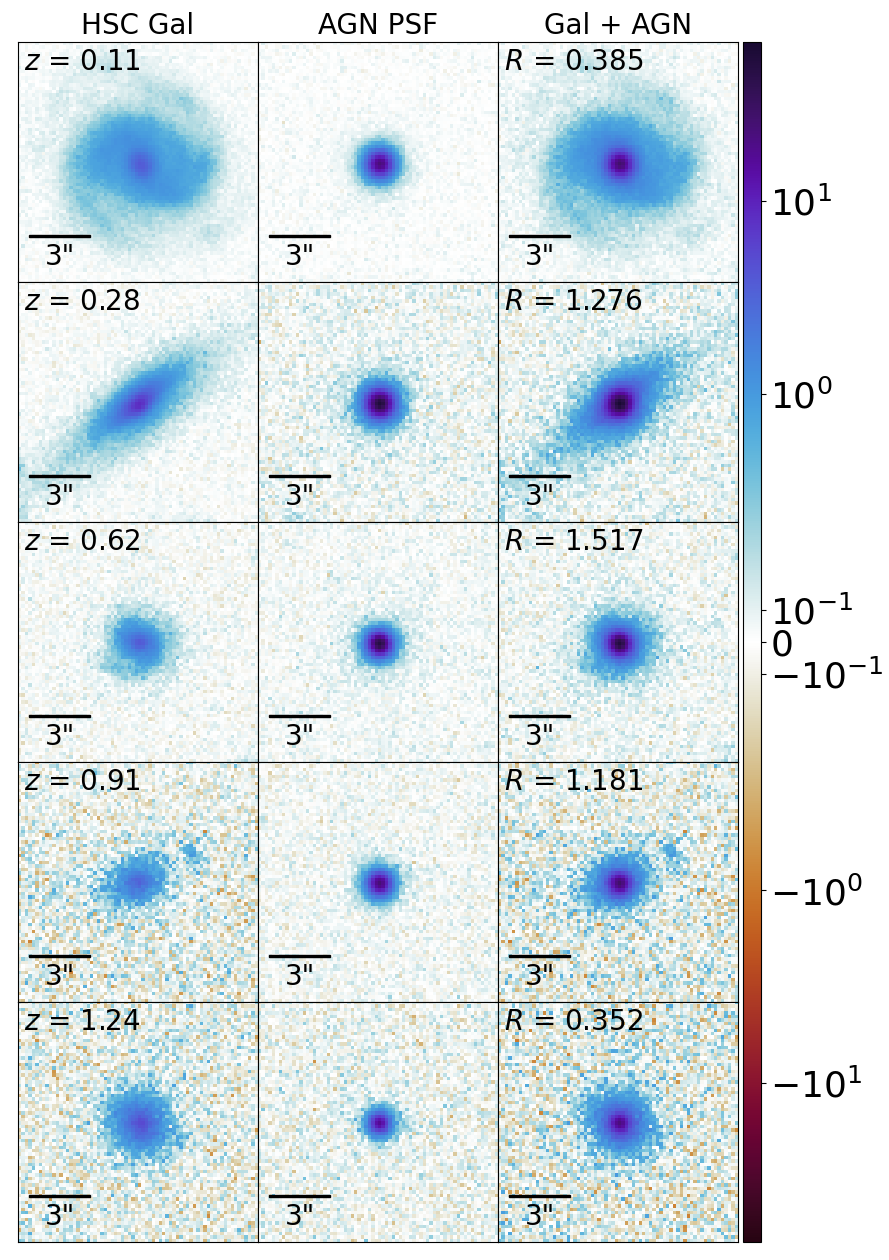}
\caption{Examples of adding artificial AGN point sources to real HSC galaxies in five redshift bins. 
{\it Left to right:} Original galaxy, AGN point source (actually a scaled stellar image), and galaxy plus AGN point source.
{\it Top to bottom:} Examples from \gbin{} ($0<z<0.25$), \rbin{} ($0.25<z<0.5$), \ibin{} ($0.5<z<0.9$), \zbin{} ($0.9<z<1.1$), and \ybin{} ($1.1<z<1.4$) redshift bins, respectively.
The photometric redshift of each example galaxy is shown in the {\it left column}.
AGN point sources were created with a wide range of AGN-to-host-galaxy contrast ratios; the value for the example is shown in the {\it right column}.
The colorbar at far right shows the flux in Analog-Digital Units (ADUs). 
We also attach a scale bar whose length equals to $3\arcsec$ to each subplot.
All subplots share the same colorbar limits.
\label{fg:agn_addition}}
\end{figure}

Before training our PSFGAN and \gampen{} models, there is some data preparation and transformation that need to be done.
Namely, we (1) create artificial AGN point sources and add them to simulated and real HSC galaxies and (2) transform the training labels associated to each galaxy.
In this subsection, we show in detail how we proceed with these two steps.

As shown in Section~\ref{sec:data} and Appendix \ref{sec:ap:galfitting}, for each redshift bin, we prepared $150,000$ simulated galaxies with known morphological parameters (from simulation) and from $\sim 12,500$ to $\sim 18,000$ labeled real HSC galaxies (number is redshift bin dependent; see Appendix \ref{sec:ap:galfitting} for details) with known morphological parameters (from GALFIT light profile fitting).
We also prepared images of $360$ local bright stars.
It is worth mentioning that while carrying out our previous experiment of combining PSFGAN with \gamornet{} \citep{2023ApJ...944..124T}, we empirically tested that using 392 local stars selected from different celestial locations would provide enough diversity in seeing conditions for our purpose of creating realistic artificial AGN point sources. 
Each AGN point source was created using 50 randomly chosen stars from this sample of 392 stars.
For this work, we choose to use 360\footnote{There is no limitation on the exact number of stars one should use, and this number can be determined empirically. In a similar scenario, one should make sure to use enough stars so that different seeing conditions are averaged out, and the exact number should also depend on the scientific question at hand.} stars to create the sample and 60 randomly selected stars to create each AGN point source.
No noticeable difference in PSFGAN performance was observed after switching to this new sample of stars for AGN creation.
All galaxy and star images are square cutouts with a common size of 185 pixels, which corresponds to $\sim 31\arcsec$ in the HSC wide survey.

In order to train PSFGAN models, we create artificial AGN point sources using these stars and add them to simulated and real HSC galaxies.
Specifically, for each simulated or real HSC galaxy, we randomly select $60$ stars from the chosen group of $360$ local bright stars.
We randomly draw an AGN-to-host-galaxy contrast ratio, $R$, from a logarithmic uniform distribution between $-1 < \log R < 0.6$ (that is, $0.1 < R < 3.981$), which is appropriate to realistic AGN host galaxies (e.g., \citealp{Gabor_2009}) and optimal for PSFGAN subtraction \citep{2018MNRAS.477.2513S}. 
We then scale the flux of each star to the desired flux $F_{desired} = R \times F_{host\_gal}$, averaged over all $60$ scaled star images\footnote{Combining them
minimizes the impact of different seeing condition.}, to create the artificial AGN point source. We then add this artificial point source to the galaxy image and repeat this process for each simulated and real HSC galaxy in all redshift bins.
For this AGN point source creation and addition procedure, Figure \ref{fg:agn_addition} shows an example in each redshift bin (with real HSC galaxies).

In addition, we transform the training labels associated with each simulated or real HSC galaxy as described in Section~5 of \citealp{2022ApJ...935..138G}.
For simulated galaxies, we use half-light radius ($R_e$) and flux ($F$) values from the simulation.
The bulge-to-total light ratio ($L_B/L_T$) is simply the bulge flux fraction for two-component galaxies.
For single-component galaxies, we set $L_B/L_T$ equal to $0$ ($1$) if its \sersic{} index is between $0.8 - 1.2$ ($3.5 - 5.0$).
For real HSC galaxies, we directly use fitted values of $L_B/L_T$, $R_e$ and $F$ from the GALFIT light profile fitting.
Since some parameters (i.e., $F$) can change over multiple orders of magnitude for different galaxies, we make appropriate transformations before using them in training. Given a set of labels:

\begin{equation}
\boldsymbol{Y_n''} = \left(L_B/L_T, R_e, F \right) ,
\label{eqt:y''}
\end{equation} 

\noindent
we first apply the logit transformation to $L_B/L_T$ and log transformations to $R_e$ and $F$:

\begin{equation}
\boldsymbol{Y_n'} = f''(\boldsymbol{Y_n''}) = \left( \log \frac{L_B/L_T}{1 - L_B/L_T}, \log R_e, \log F \right) ,
\label{eqt:y'}
\end{equation} 
where $f''$ stands for logit\footnote{$\text{logit}(x)=\ln{\frac{x}{1-x}}$}/log transformation. We then apply the standard scaling (calibrated on the training data) independently on each one of $L_B/L_T$, $R_e$, and $F$ so that the parameter mean across the training data is shifted to zero and the parameter variance equals one:

\begin{equation}
\boldsymbol{Y_n} = f'(\boldsymbol{Y_n'}) ,
\label{eqt:y}
\end{equation} 
where $f'$ stands for the standard scaling. Scaled parameters can now be used safely in \gampen{} training, as contributions from different parameters are now balanced and the network is more robust against numerical instabilities.

After training our \gampen{} models and applying them on test sets, for each sample $\boldsymbol{\hat{Y}_{n,t}}$ (drawn from the corresponding \gampen{} predicted multivariate normal distribution $\mathcal{N}\left(\hat{\boldsymbol{\mu}}_{n,t},\boldsymbol{\hat{\Sigma}}_{n,t}\right)$), we unscale each of its parameters with respect to the training-data-based calibration:

\begin{equation}
\boldsymbol{\hat{Y'}_{n,t}} = f'^{-1}(\boldsymbol{\hat{Y}_{n,t}}) = \left( \log \frac{L_B/L_T}{1 - L_B/L_T}, \log R_e, \log F \right) ,
\label{eqt:y'_i}
\end{equation} 
and then perform the inverse transformation:
\begin{equation}
\boldsymbol{\hat{Y''}_{n,t}} = f''^{-1}(\boldsymbol{\hat{Y'}_{n,t}}) = \left(L_B/L_T, R_e, F \right) .
\label{eqt:y''_i}
\end{equation} 

\noindent
Note that this procedure also ensures that the predicted values always conform to physically meaningful ranges ($0 \leq L_B/L_T \leq 1$; $R_e > 0$; $F > 0$).

\subsection{Initial Training of PSFGAN and \gampen{} Models on Simulated Galaxies} \label{sec:init_training}

In this subsection, we present details on how we train the initial PSFGAN and \gampen{} models with simulated data.

For each of the five redshift bins, we train a separate model of PSFGAN and a separate model of \gampen{} (we use images from the corresponding band for both models).
Before training, we split the $150,000$ simulated galaxies (Section~\ref{sec:sim_gal}) in each bin into five smaller subsets: PSFGAN training set ($9,000$ galaxies), PSFGAN validation set ($1,000$ galaxies), training set of \gampen{} ($98,000$ galaxies), validation set of \gampen{} ($21,000$ galaxies), and the common test set ($21,000$ galaxies).
In doing so, we make sure that the distribution of $L_B/L_T$ is balanced in each of the five subsets (by adding a constraint when splitting the data).
Besides this, the split is completely random with respect to other parameters, including the ratio of double to single-component galaxies.
It is also worth mentioning that we intentionally assign the majority of galaxies to train \gampen{} from scratch. This is because, typically, discriminative models like \gampen{} require a much larger training set compared to generative models (like PSFGAN), which can be adequately trained with a much smaller training set.

We first train our PSFGAN model using the training and validation sets of PSFGAN.
As mentioned in Section~\ref{sec:prep_trans}, we added an artificial AGN point source to each simulated galaxy.
We now train a PSFGAN model to remove this added AGN point source and to recover the host galaxy.
All images in the training and validation sets of PSFGAN are square cutouts of 185 pixels ($\sim 31\arcsec$).
By design \citep{2018MNRAS.477.2513S}, PSFGAN has the following loss function structure (loss functions followed by abbreviation definitions):

\begin{equation}
\begin{split}
\label{eqt:1}
    L_{gen} &= - \log (\sig (D(ctcp(X^{gen}), ctcp(X^{cond})))) \\
    &+ \frac{\lambda P_{att}}{N^2}\sum_{\textrm{all pixels}}|X^{ori} - X^{gen}|_{ij} \\
    &+ \frac{\lambda}{{S_{att}}^2}\sum_{\textrm{all pixels}}|ctcp(X^{ori}) - ctcp(X^{gen})|_{ij} ; \\
    L_{dsc} &=  - \log (\sig (D(ctcp(X^{ori}), ctcp(X^{cond})))) \\
    &- \log (1 - \sig (D(ctcp(X^{gen}), ctcp(X^{cond})))) ;\\
    &\sig(x) = \frac{e^{x}}{1+e^{x}} , \\ 
\end{split}
\end{equation}

\noindent
where $L_{dsc}$ stands for the discriminator loss and $L_{gen}$ stands for the generator loss. \\

When dealing with realistic CCD images, there are usually multiple galaxies shown on the same cutout.
However, the task of AGN point source removal is carried out only for the galaxy at the image center.
For this reason, we use central-cropped images (which only contains pixels within a central attention window of the uncropped image) for the discrimanator inputs (see the next paragraph).
The size (in pixels) of this attention window, $S_{att}$, is a hyperparameter, while $N$ is the image size in pixels (always fixed to $185$).
$ctcp(X)$ denotes the central-cropped image of an image $X$. For a square image $X$ of $N$ pixels, its indices $i$ and $j$ can range from $0$ to $N-1$. We select all pixels $i$ and $j$ that range from $(N-1)/2-(S_{att}/2)+1$ to $(N-1)/2+(S_{att}/2)$ to create the central-cropped image $ctcp(X)$, which is itself a square image of $S_{att}$ pixels.

As mentioned in Section~\ref{sec:intro_psfgan}, the discriminator always maps an input of 
(input galaxy, host galaxy + AGN point source) to a scalar output.
Here, the input galaxy ($ctcp(X^{input})$) can either be a real host galaxy ($ctcp(X^{ori})$) or a fake host galaxy from the generator ($ctcp(X^{gen})$).
Likewise, we use $ctcp(X^{cond})$ to denote the host galaxy + AGN point source and $D(ctcp(X^{input}), ctcp(X^{cond}))$ to denote the discriminator's output, which represents the probability that this input galaxy is a real one according to the discriminator.
Therefore, to train PSFGAN's discriminator, we want to maximize $D(ctcp(X^{ori}), ctcp(X^{cond}))$ and to minimize $D(ctcp(X^{gen}), ctcp(X^{cond}))$.
This explains why $L_{dsc}$ takes the shown form. 
The sigmoid function, $\sig(x)=e^x/(1+e^x)$, was added to ensure that both $\sig(D(ctcp(X^{input}), ctcp(X^{cond})))$ and $1-\sig(D(ctcp(X^{input}), ctcp(X^{cond})))$ are larger than zero so values of the logarithmic function are always valid.
Likewise, to train PSFGAN's generator, we want to maximize $D(ctcp(X^{gen}), ctcp(X^{cond}))$, which determines the first term in $L_{gen}$.
\\

The second and third terms in $L_{gen}$ are the (normalized) averaged values of the absolute difference between $X^{ori}$ and $X^{gen}$ and between $ctcp(X^{ori})$ and $ctcp(X^{gen})$, respectively.
Regularization parameters ($\lambda=100$ and $P_{att}=0.05$) are inherited from the original PSFGAN paper \citep{2018MNRAS.477.2513S} and they work well with our experiments.
These terms added a constraint that $X^{gen}$ and $X^{ori}$ should be similar and the network ``cares about'' this similarity $20$ times more in the central attention window than in the remaining part of the image.
At last, to minimize the loss function, we use Adam, a well-known first-order gradient-based optimizer \citep{2014arXiv1412.6980K}.

\begin{deluxetable*}{cccccc}[htbp]
\tablecaption{Redshift-bin-dependent Hyper Parameters for Initial Training of PSFGAN (\textit{Top}) and \gampen{} (\textit{Bottom}) on Simulated Galaxies}
\label{tb:init_para}
\tablecolumns{6}
\tablehead{
\colhead{} & \colhead{\gBin{}} & \colhead{\rBin{}} & \colhead{\iBin{}} & \colhead{\zBin{}} & \colhead{\yBin{}}\\
\colhead{} & \colhead{(\gb{} band)} & \colhead{(\rb{} band)} & \colhead{(\ib{} band)} & \colhead{(\zb{} band)} & \colhead{(\yb{} band)} 
} 
\startdata
    \hline
    $S_{att}$\tablenotemark{a} & $22$ & $16$ & $14$ & $12$ & $12$ \\
    $LR$\tablenotemark{b} & $5\times10^{-5}$ & $1.5\times10^{-5}$ & $2\times10^{-5}$ & $1\times10^{-5}$ & $1\times10^{-5}$ \\
    $n_{epoch}$\tablenotemark{c} & $40$ & $40$ & $40$ & $50$ & $50$ \\
    \hline
    \texttt{Batch size} & $32$ & $8$ & $8$ & $16$ & $16$ \\
    \texttt{Learning rate} & $5\times10^{-6}$ & $5\times10^{-6}$ & $5\times10^{-6}$ & $1\times10^{-6}$ & $5\times10^{-7}$ \\
    \texttt{Momentum} & $0.99$ & $0.95$ & $0.9$ & $0.9$ & $0.9$ \\
    \texttt{L2 regularization} & $1\times10^{-4}$ & $1\times10^{-2}$ & $1\times10^{-5}$ & $1\times10^{-4}$ & $1\times10^{-5}$ \\
    \texttt{Dropout rate} & $7\times10^{-4}$ & $5\times10^{-4}$ & $4\times10^{-4}$ & $3\times10^{-4}$ & $3\times10^{-4}$ \\
    \hline
    \hline
\enddata
\tablenotetext{a}{The length of each side, in pixels, of the square attention window at the image center.}
\tablenotetext{b}{Governs the parameter increment at each learning step.}
\tablenotetext{c}{How many times the entire training set of PSFGAN is used for training the current model (neural networks are usually trained for multiple epochs).}
\end{deluxetable*}

Prior to feeding images to PSFGAN, we perform a normalization on all input images (both images of the original galaxy and images of the original + AGN), and we perform the inverse operation on all PSFGAN-generated images before using them for other purposes (e.g., training \gampen{}).
Every pixel flux value, $X_{ij}$, is transformed under an ${\rm arcsinh}$ stretch function: 
\begin{equation}
\label{eqt:2}
    X^{\prime}_{ij} = \frac{{\rm arcsinh} (A \times X_{ij})}
    {{\rm arcsinh} (A \times X_{max})}.
\end{equation}

\noindent
Hyperparameters present in the stretch function include:
\begin{itemize}
    \item $X_{max}$ (maximum pixel flux value for the entire set of images to be normalized);
    \item $A=50$ (stretch factor).
\end{itemize}

\noindent
We found this normalization process very helpful, particularly in cases where the pixel flux values span a few orders of magnitude.

In addition, there are a few other hyperparameters in PSFGAN training:
\begin{itemize}
    \item $LR$ (learning rate);
    \item $n_{epoch}$ (number of epochs).
\end{itemize}

\noindent
Hyperparameters that take the same value across all redshift bins have their values reported when we introduce them.
Three hyperparameters are redshift-bin dependent, and their values are shown in Table \ref{tb:init_para} (\textit{top panel}).
We determine hyperparameters to be the values that give the best results on the validation set of PSFGAN, for each redshift bin.

Once the PSFGAN model in each redshift bin is properly trained, we apply it to the training and validation sets of \gampen{} and the common test set, to remove the added AGN point source and to recover the host galaxy.
We then used the recovered galaxies in the training and validation sets of \gampen{} to train our \gampen{} model.
We crop images to smaller sizes\footnote{Although the STN module can determine the optimal cropping size, this initial cropping speeds up the converging process in training, particularly for higher redshift bins in which galaxy sizes are quite small. On another hand, we know that the image size after this initial cropping is still larger than 10 times the upper bound of galaxy's Kron radius in each redshift bin (see the last panel in Figure \ref{fg:hsc_gal_pre}), so no essential information (for the galaxy at image center) is lost.} that are appropriate for galaxy sizes in the corresponding redshift bin: all images in the training and validation sets of \gampen{} and the common test set are cropped to a square size of 179, 143, and 95 pixels (corresponding to $\sim 30\arcsec$, $\sim 24\arcsec$, and $\sim 16\arcsec$ in the HSC Wide survey), for redshift bin \gBin, \rBin, and the rest three redshift bins, respectively.
We also use the \texttt{arcsinh} function to normalize each image before feeding it to the \gampen{}.
Also, recall that in Section~\ref{sec:prep_trans}, we have transformed and scaled the training labels associated with each galaxy.
We use these scaled labels when calculating the loss at each training step.

Specifically, for each redshift bin, we train a \gampen{} model from scratch by minimizing its loss function using Stochastic Gradient Descent \citep{2016arXiv160904747R}.
Detailed structures of \gampen{}, including its loss function, are comprehensively discussed in \citealp{2022ApJ...935..138G}, and we refer interested readers to it.
The \href{https://gampen.readthedocs.io/en/latest/Using_GaMPEN.html#}{Tutorial Page of \gampen{}} also contains a description for each of all tuneable parameters; here we list a few key hyperparameters that are vital in training \gampen{} models:

\begin{itemize}
    \item \texttt{Batch size}: the number of images processed before weights and biases are updated;
    \item $\texttt{Epochs} = 40$: the analogous hyper-parameter to $n_{epoch}$ discussed in PSFGAN training;
    \item \texttt{Learning rate}: the analogous hyper-parameter to $LR$ discussed in PSFGAN training;
    \item \texttt{Momentum}: an acceleration factor used for faster convergence;
    \item \texttt{L2 regularization}: the degree to which larger weight values are penalized; it has a similar role to the regularization parameter discussed in PSFGAN training (though one should bear in mind their roles are not exactly the same);
    \item \texttt{Dropout rate}: an independent probability under which each neuron is dropped during each feed-forward pass of an input; see Section~\ref{sec:intro_cnn} for details.
\end{itemize}

Apart from \texttt{Epochs}, which was fixed to $40$ for all redshift bins, we tried a few values of \texttt{batch size}, \texttt{learning rate}, \texttt{momentum}, \texttt{L2 regularization} and \texttt{dropout rate} in each bin. 
To summarize: $8$, $16$, $32$, $64$ for \texttt{batch size}; $5\times10^{-6}$, $1\times10^{-6}$, $5\times10^{-7}$, $1\times10^{-7}$ for \texttt{learning rate}; $0.8$, $0.9$, $0.95$, $0.99$ for \texttt{momentum}; $1\times10^{-2}$, $1\times10^{-3}$, $1\times10^{-4}$, $1\times10^{-5}$ for \texttt{L2 regularization}; $7\times10^{-4}$, $5\times10^{-4}$, $4\times10^{-4}$, $3\times10^{-4}$ for \texttt{dropout rate}.
We determine these hyperparameters to be the values that give the best results on the validation set of \gampen{}, for each redshift bin.
The selected values of these parameters are shown in Table \ref{tb:init_para} (\textit{bottom panel}).
We then apply the trained \gampen{} model on the common test set to make sure it yields expected results.
It is worth mentioning that, instead of analyzing the application results of our initial \gampen{} model on the common test set of simulated data, we choose to comprehensively analyze the results of our final \gampen{} model applied on the real test data---for the latter are the actual models to be used to study active galaxy morphological parameters in the HSC Wide survey (see Section~\ref{sec:perfo} for details).
We discuss the details of the transfer learning step in the next section.

\subsection{Transfer Learning of PSFGAN and \gampen{} Models on Real HSC Galaxies} \label{sec:trans_learning}

\begin{deluxetable*}{cccccc}[htbp]
\tablecaption{Redshift-bin-dependent Hyper Parameters for Transfer Learning of PSFGAN (\textit{Top}) and \gampen{} (\textit{Bottom}) on HSC Galaxies}
\label{tb:trans_para}
\tablecolumns{6}
\tablehead{
\colhead{} & \colhead{\gBin{}} & \colhead{\rBin{}} & \colhead{\iBin{}} & \colhead{\zBin{}} & \colhead{\yBin{}}\\
\colhead{} & \colhead{(\gb{} band)} & \colhead{(\rb{} band)} & \colhead{(\ib{} band)} & \colhead{(\zb{} band)} & \colhead{(\yb{} band)} 
} 
\startdata
    \hline
    $S_{att}$ & $22$ & $16$ & $14$ & $12$ & $12$ \\
    $LR$ & $5\times10^{-5}$ & $1.5\times10^{-5}$ & $2\times10^{-5}$ & $8\times10^{-6}$ & $8\times10^{-6}$ \\
    $n_{epoch}$ & $20$ & $20$ & $20$ & $40$ & $40$ \\
    \hline
    \texttt{Batch size} & $8$ & $8$ & $8$ & $16$ & $32$ \\
    \texttt{Learning rate} & $5\times10^{-7}$ & $1\times10^{-6}$ & $1\times10^{-6}$ & $1\times10^{-6}$ & $1\times10^{-6}$ \\
    \texttt{Momentum} & $0.99$ & $0.95$ & $0.95$ & $0.9$ & $0.99$ \\
    \texttt{L2 regularization} & $1\times10^{-3}$ & $1\times10^{-4}$ & $1\times10^{-5}$ & $1\times10^{-5}$ & $1\times10^{-5}$ \\
    \texttt{Dropout rate} & $4\times10^{-4}$ & $2\times10^{-4}$ & $2\times10^{-4}$ & $1.5\times10^{-4}$ & $1.5\times10^{-4}$ \\
    \hline
    \hline
\enddata
\end{deluxetable*}

We now move to the transfer learning phase, using images of real HSC galaxies (with added AGN point sources).
We train PSFGAN models from scratch and use them to generate recovered images of host galaxies.
We then take \gampen{} models previously trained on simulated data and fine-tune their parameters with respect to these recovered galaxies.

First of all, as in the initial training phase, we split the entire set of real HSC galaxies into five subsets.
From GALFIT light profile fitting (see Appendix \ref{sec:ap:galfitting}) we have $17789$, $15542$, $13994$, $13706$, and $12519$ labeled HSC galaxies in the redshift bins \gbin{}, \rbin{}, \ibin{}, \zbin{}, and \ybin{}, respectively (Figure \ref{fg:hsc_gal_para}).
In each redshift bin, we randomly select $4,500$ galaxies to create the training set of PSFGAN, and another $500$ galaxies to create the validation set of PSFGAN.
The rest of the galaxies are used to create the other three subsets: $70 \%$ of them are assigned to the training set of \gampen{}, $5 \%$ of them are assigned to the validation set of \gampen{}, and the last $25 \%$ of them are assigned to the common test set.
When making these splits, we make sure that the distribution of $L_B/L_T$ is balanced in each of the five subsets.
The split is completely random with respect to other parameters.

For each redshift bin, we train a separate model of PSFGAN from scratch, using the corresponding training and validation sets of PSFGAN, both of which contain square cutouts of 185 pixels ($\sim 31\arcsec$) of the original galaxy and the original + AGN.
The techniques used to train PSFGAN models with real data are exactly the same as those we used in the previous subsection (i.e., with simulated data, see Section~\ref{sec:init_training}).
Thus redundant descriptions such as loss functions, image normalization, etc. are skipped in this subsection.
All redshift-bin-independent values (e.g., $\lambda$, $P_{att}$) are kept the same as for the simulated data; we determine redshift-bin-dependent hyperparameters based on model performance on the validation set of PSFGAN, for each redshift bin.
Their values are shown in Table \ref{tb:trans_para} (\textit{top panel}).

Next, we apply the trained PSFGAN model on the training and validation sets of \gampen{} and the common test set, for each redshift bin, to create images of recovered host galaxies.
Before using these recovered galaxy images, we notice that for a generic redshift bin, the size of the training set of \gampen{} for real HSC galaxies is about an order of magnitude smaller than the size of the same subset for simulated galaxies.
Moreover, the distribution of magnitude for the entire set of real HSC galaxies is heavily biased towards fainter galaxies (see Figure \ref{fg:hsc_gal_para}), as is the distribution of magnitude for the training set of \gampen{} (for we split the entire set into five subsets randomly with respect to magnitude).
Since \gampen{} is a discriminative model, it generally requires more images in the training set than for generative models (e.g., PSFGAN)---a feature we mentioned in the data splitting step of the previous subsection.
In addition to that, as we have extensively tested in our previous works \citep{2020ApJ...895..112G, Ghosh_2023}, it is important to use a training set for \gampen{} that spans evenly across magnitude (despite that not being the case in the real universe), in order to achieve a similar model performance on both brighter and fainter galaxies.
For these reasons, we perform a data augmentation step on the training set of \gampen{} in each redshift bin, before proceeding to train the \gampen{} models.
Details of this data augmentation step are summarized below.

We start with the \gbin{} redshift bin, in which we divide the training set of \gampen{} into three smaller subsets by magnitude---one with $m<19$, another one with $19<m<21$, and the third with $m>21$.
For each galaxy in these three subsets, we create a few of its copies by (1) a rotation of random angle drawn from a uniform distribution between $0^{\circ}$ and $360^{\circ}$, (2) a possible reflection along the x axis ($50\%$ chance), and (3) another possible reflection along the y axis (also $50\%$ chance; all probabilities are mutually independent). 
Since there are far more faint galaxies than bright galaxies in the original training set of \gampen{}, we chose to create more copies for the brighter ones than the fainter ones.
After the three subsets are all augmented, we combine them into one (augmented) training set, which now has an even distribution of magnitude (i.e., there are a similar number of galaxies within each of the $m<19$, $19<m<21$ and $m>21$ magnitude ranges).
We repeat this process in the other four redshift bins with slightly different magnitude boundaries: $m<20$, $20<m<21.5$ and $m>21.5$ for the \rbin{} redshift bin, $m<20.5$, $20.5<m<22$ and $m>22$ for the \ibin{} redshift bin, and $m<21$, $21<m<22$ and $m>22$ for the \zbin{} and \ybin{} redshift bins.
Once we arrive at a new training set of \gampen{} (which has an even distribution on magnitude) in each of the five redshift bins, we again augment each one of them so that we end up with $\sim 70,000$ galaxies at the end.
This step guarantees that in each redshift bin: (1) there are enough galaxies in the training set of \gampen{}, and (2) the distribution of galaxy magnitude in the training set of \gampen{} is even.

In each redshift bin, using PSFGAN-recovered galaxies in the augmented training set of \gampen{} as well as the corresponding validation set, we fine-tune parameters of the \gampen{} model previously trained with simulated data.
As in the previous subsection, we crop images to smaller sizes that are appropriate for galaxy sizes in the corresponding redshift bin: images in the training and validation sets of \gampen{} and the common test set are cropped to a square size of 179, 143, and 95 pixels (corresponding to $\sim 30\arcsec$, $\sim 24\arcsec$, and $\sim 16\arcsec$), for redshift bin \gBin, \rBin, and the rest three redshift bins, respectively.
We again use the \texttt{arcsinh} function for image normalization, use transformed and scaled training labels associated to each galaxy for loss function calculation, and use Stochastic Gradient Descent for loss function minimization.

We determine hyperparameters when fine-tuning each \gampen{} model in the transfer learning phase by analyzing its performance on the validation set of \gampen{}, for each redshift bin.
Choices of hyperparameters are: $4$, $8$, $16$, $32$, $64$ for \texttt{batch size}; $5\times10^{-6}$, $1\times10^{-6}$, $5\times10^{-7}$, $1\times10^{-7}$ for \texttt{learning rate}; $0.8$, $0.9$, $0.95$, $0.99$ for \texttt{momentum}; $1\times10^{-2}$, $1\times10^{-3}$, $1\times10^{-4}$, $1\times10^{-5}$ for \texttt{L2 regularization}; $4\times10^{-4}$, $2\times10^{-4}$, $1.5\times10^{-4}$, $1\times10^{-4}$ for \texttt{dropout rate}.
\texttt{Epochs} is again fixed to $40$ for all redshift bins.
Chosen values of hyperparameters are shown in Table \ref{tb:trans_para} (\textit{bottom panel}).

After fine-tuning parameters of the \gampen{} models, in each redshift bin, we apply the final (parameter-fine-tuned) \gampen{} model on PSFGAN-recovered galaxies in the common test set\footnote{These final PSFGAN and \gampen{} models are the ones we aim to use in studying active galaxies in HSC Wide survey.}.
In doing so, each recovered galaxy image is passed to the final \gampen{} model for $T=1,000$ times, and we aggregate the values of $\boldsymbol{\hat{Y''}_{n,t}}$ from each feed-forward pass to make the predicted posterior distributions for $L_B/L_T$, $R_e$ and $F$.
We analyze these results in the next section.

\section{Model Performance} \label{sec:perfo}
In this section, we analyze the performance of our trained models on the common test set (of real HSC galaxies) in all five redshift bins.
We start by inspecting the performance of the model on some selected examples (Section~\ref{sec:perfo_idvl}).
We then move on to model accuracies (Section~\ref{sec:perfo_acrc}) and uncertainties (Section~\ref{sec:perfo_uctt}), both across the entire common test set.

\subsection{Performance on Individual Examples} \label{sec:perfo_idvl}

\begin{figure}
\figurenum{7}
\gridline{\fig{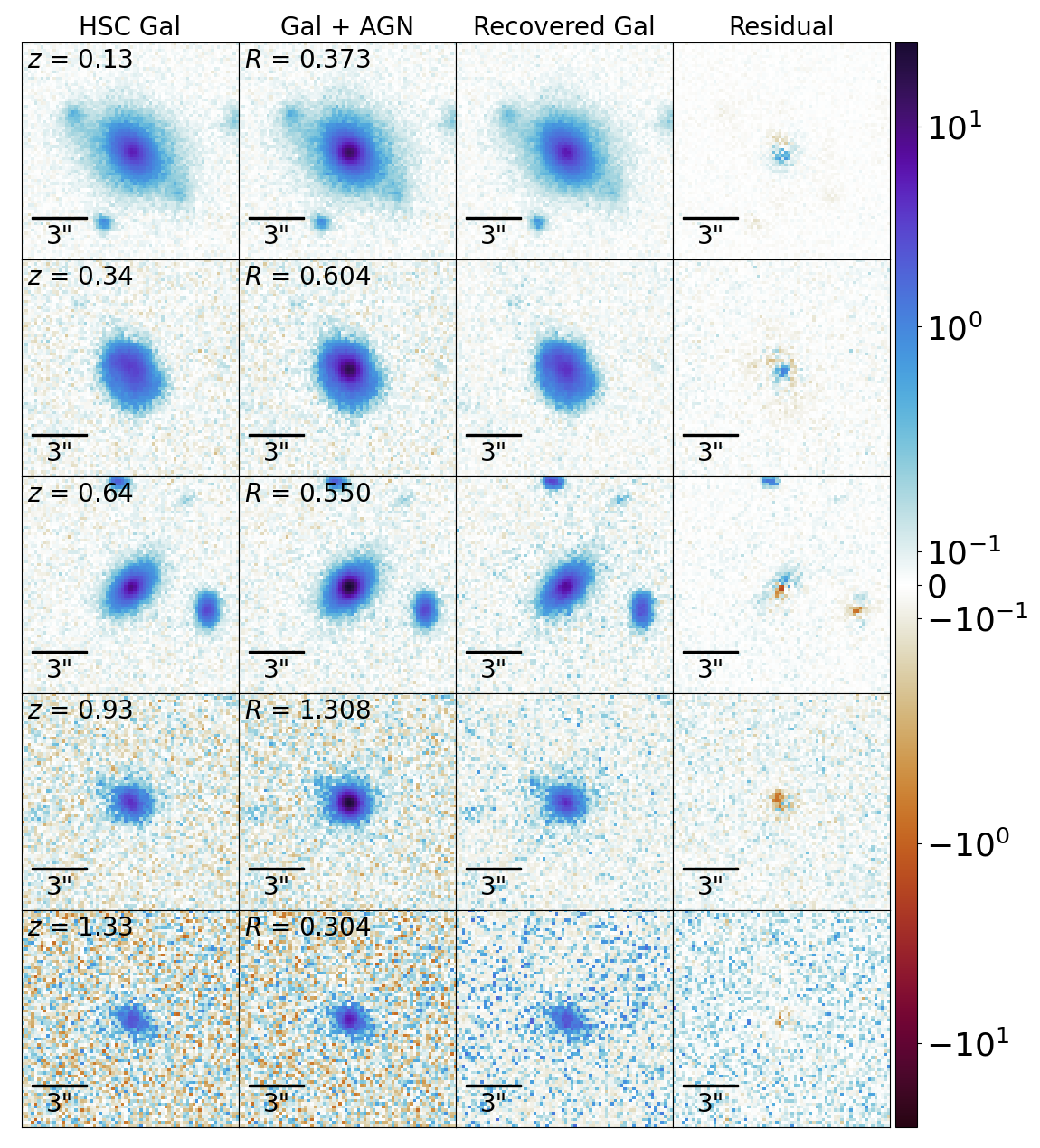}{0.52\textwidth}{}}
\vspace*{-0.4cm}
\caption{
Examples showing PSFGAN performance on images of real HSC galaxy + artificial AGN point source in all five redshift bins. 
Each row corresponds to an example chosen from an unique redshift bin 
({\it top to bottom}): 
\gbin{}, \rbin{}, \ibin{}, \zbin{}, and \ybin{}.
{\it Left to right:}
original galaxy (from the HSC Wide survey), 
original + AGN point source, 
galaxy recovered by PSFGAN, 
and the residual image (recovered galaxy minus original galaxy). 
Labels in the boxes represent photometric redshifts ({\it z; first column}) and AGN-to-host-galaxy flux contrast ratios ({\it R; second column}).
The small residuals show that PSFGAN works well to recover AGN host galaxies.
In-subplot scale bars and the common colorbar are created using the same format as in Fig.\,\ref{fg:agn_addition}.
\label{fg:psf_gmp_a}}
\end{figure}

\begin{figure}
\figurenum{8}
\gridline{\fig{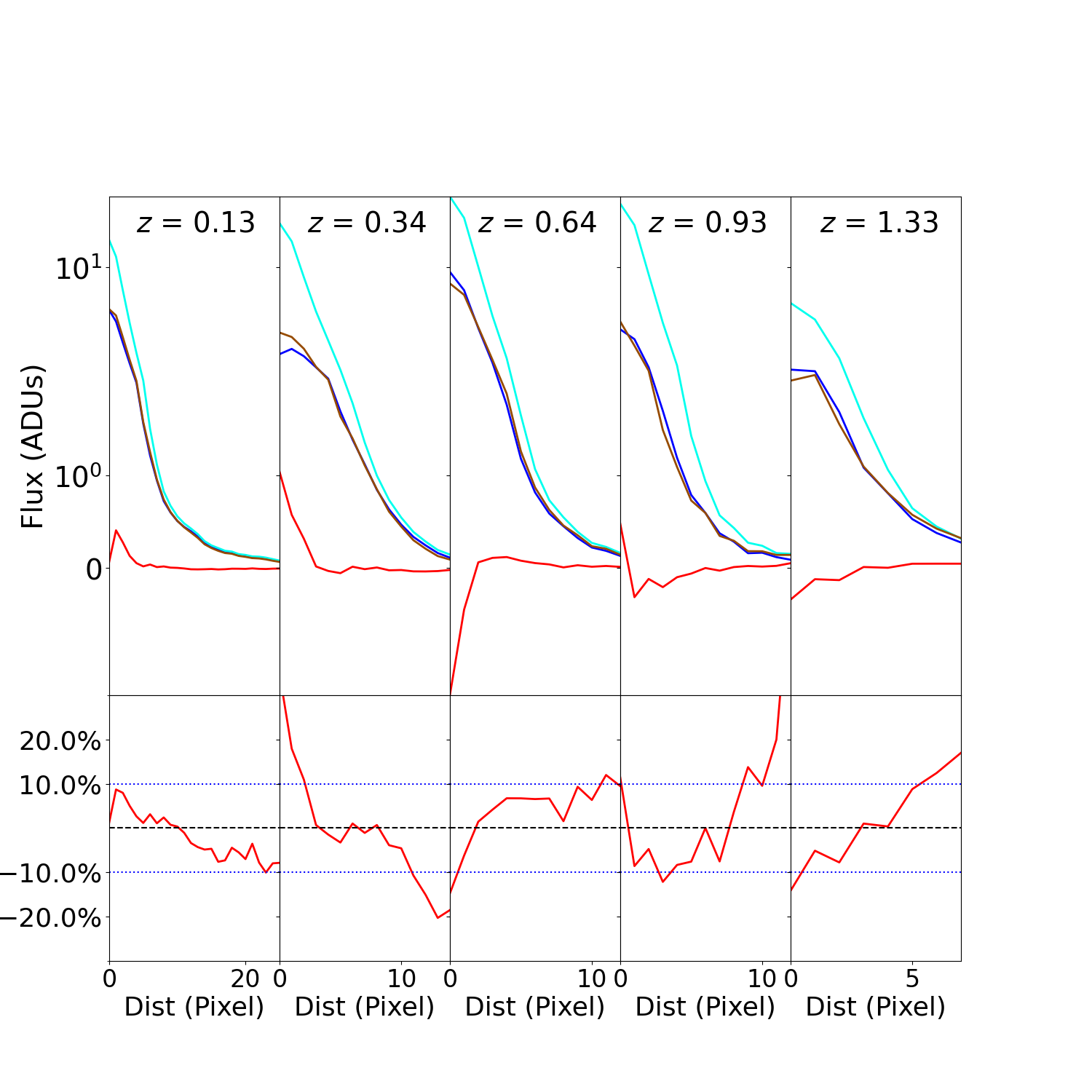}{0.5\textwidth}{}}
\vspace*{-0.4cm}
\caption{
Radial profiles ({\it top}) of the five examples from Fig.\,\ref{fg:psf_gmp_a}, for the original galaxy ({\it blue curve}), original + AGN point source ({\it cyan curve}), recovered galaxy ({\it brown curve}), and residual ({\it red curve}, defined as original minus recovered flux). 
Photometric redshift is shown at the top of each subplot.
Bottom panels show the residual ({\it red curve}) expressed as the percentage of the flux of the original galaxy. 
At the majority of radii, the residuals are within 10 percentage of the corresponding original galaxy flux.
\label{fg:psf_gmp_b}}
\end{figure}

\begin{figure*}[htb]
\figurenum{9}
\centering
\includegraphics[width=0.9\textwidth]{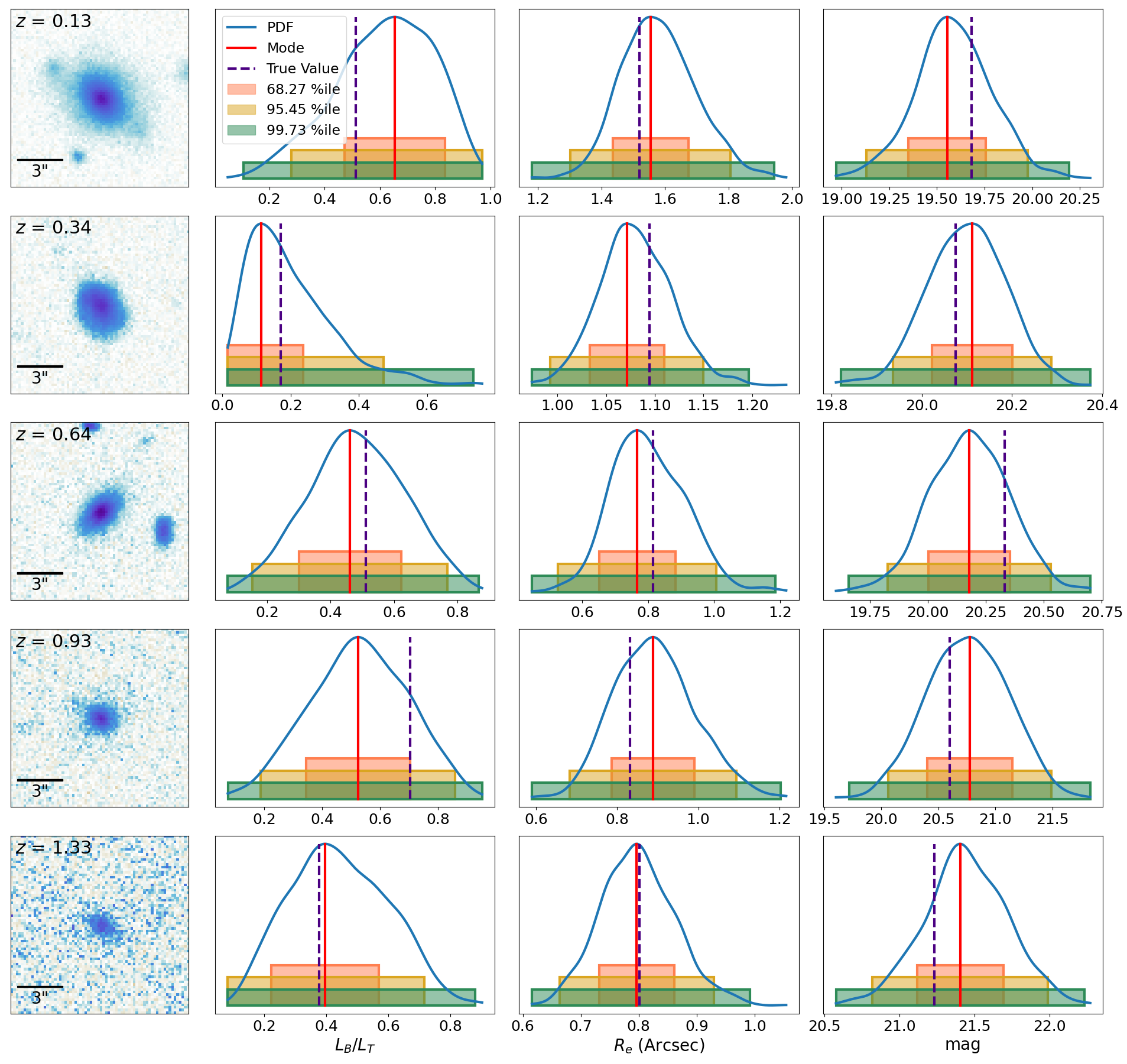}
\caption{
Examples showing \gampen{} performance: predicted posterior distributions of the three morphological parameters for the five examples from Fig.\,\ref{fg:psf_gmp_a}.
The solid blue curves show the predicted probability distribution functions (calculated by Kernel Density Estimation), and the solid red lines show the mode (defined as the most probable value of each morphological parameter).
We also show 1, 2, and 3 $\sigma$ confidence intervals (the orange, yellow, and green shaded boxes, respectively).
In the vast majority of cases, the ``true'' value ({\it purple dashed line}, determined from GALFIT light profile fitting) lies within the 1 $\sigma$ confidence interval.
For convenience, we attach the image of PSFGAN-recovered galaxy (with redshift label and scale bar) of each example at the beginning of each row.
\label{fg:psf_gmp_c}}
\end{figure*}

After we have finished training our PSFGAN and \gampen{} models, we apply both to the common test set.
Specifically, we first use the trained PSFGAN model to remove added AGN point sources and to recover host galaxies, then send these recovered galaxies to the trained \gampen{} models in order to estimate posterior probability distributions of the three morphological parameters ($L_B/L_T$, $R_e$ and $F$).
Note that in each redshift bin we have trained a separate PSFGAN model and a separate \gampen{} model---likewise, the above procedure is carried out separately in each bin as well.

We begin by analyzing the performance of our PSFGAN models.
We randomly select an example from the common test set of each redshift bin; Figure \ref{fg:psf_gmp_a} shows images of the original galaxy, original + AGN point source, recovered galaxy and the residual (recovered minus original), for each example.
As expected, our trained PSFGAN models are capable of removing a significant amount of light from the added AGN point source and only leaving a fractional pattern at each image center.
We further analyze this residual pattern by plotting radial profiles of (again) the original galaxy, original + AGN point source, recovered galaxy and the residual, for each of the chosen examples, in Figure \ref{fg:psf_gmp_b}.
At the bottom of Figure \ref{fg:psf_gmp_b}, we show the fractional radial profile of the residual (defined as the residual profile divided by the original profile), which is within $10\%$ at the majority of radii, thus validating that our PSFGAN models work well on these examples.

We then analyze the performance of our \gampen{} models.
By design (Section~\ref{sec:overv_appro} and Section~\ref{sec:trans_learning}), each recovered image is fed to \gampen{} for $T=1,000$ times.
Each time the input is passed to a slightly different network, giving a unique multivariate normal distribution, from which we draw a sample.
These samples are inversely transformed and then aggregated to make posterior probability distributions of $L_B/L_T$, $R_e$ and $F$.
The inverse transformation ensures that our predicted values are within physically reasonable ranges ($0 \leq L_B/L_T \leq 1$; $R_e > 0$; $F > 0$).

The above procedure generates predicted posterior probability distributions of morphological parameters\footnote{From now on, when characterizing galaxy brightness, we use apparent magnitude, $mag$, in lieu of flux, $F$, for an easier comparison with other works.} for recovered galaxies in the common test set.
Figure \ref{fg:psf_gmp_c} shows the details for each of our chosen examples: for each example/parameter, we use kernel density estimation (KDE) to calculate the predicted probability distribution function (solid blue curve).
The highest value of each probability distribution function (mode) is what we define as the ``predicted value'' from \gampen{} (solid red line).
Moreover, the confidence intervals 1, 2 and 3 $\sigma$ are also shown.
In almost all cases (14 out of 15), the true value (purple dashed line), which comes from the fitting of the GALFIT light profile, lies within the 1 $\sigma$ confidence interval.
Overall, after PSFGAN removes the added AGN point source, \gampen{} shows robust performance in predicting the distributions of $L_B/L_T$, $R_e$, and $mag$ in these examples without any noticeable confusion caused by the AGN point source.

\subsection{Evaluating Model Accuracies} \label{sec:perfo_acrc}

\begin{figure}
\figurenum{10}
\gridline{\fig{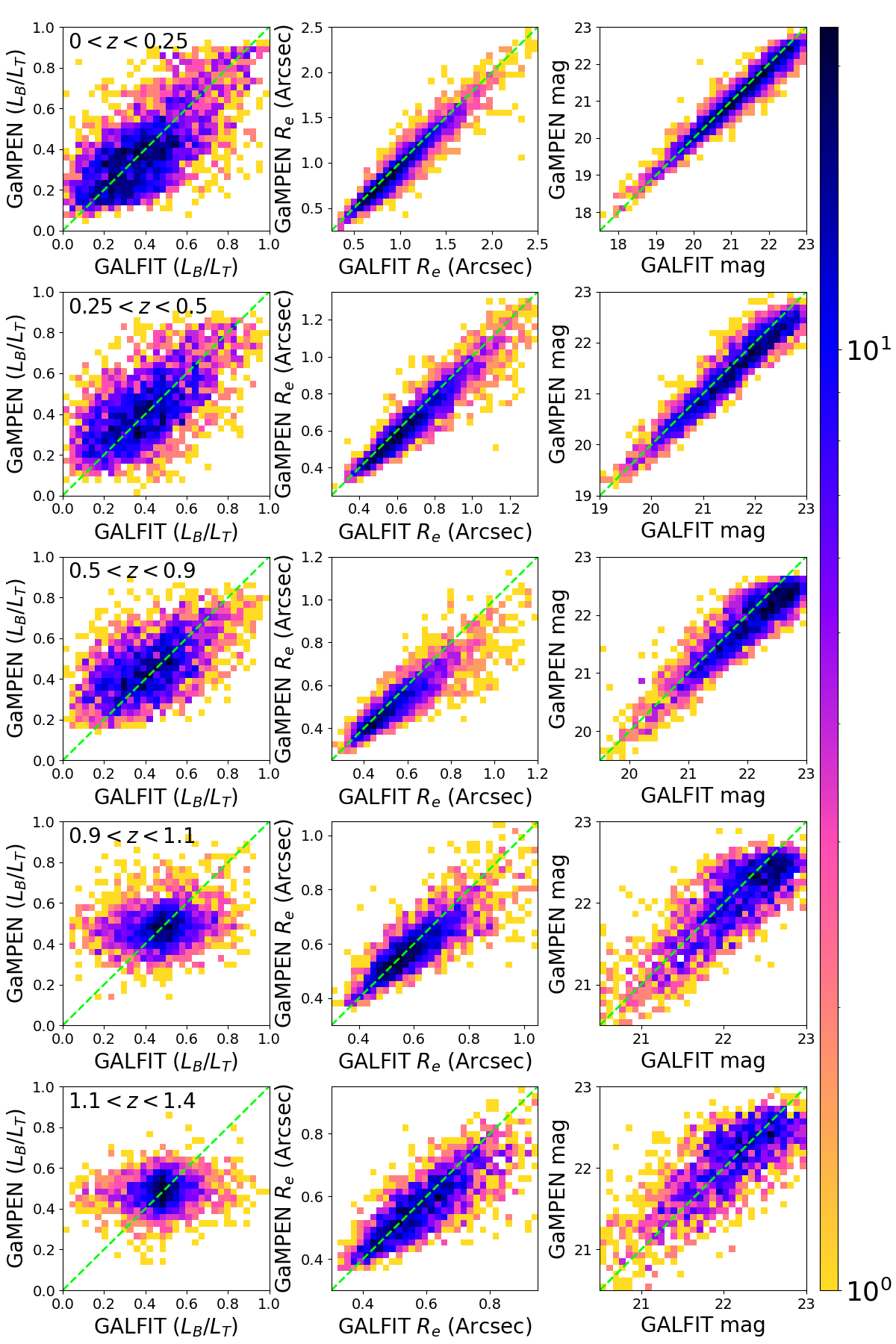}{0.52\textwidth}{}}
\vspace*{-0.4cm}
\caption{
Comparison of predicted values (\gampen{}) and true values (GALFIT) for galaxies in the common test set 
for each of the three morphological parameters, and for each of the five redshift bins.
We color each histogram using a common logarithmic colorbar (shown on the right) to better visualize the number of galaxies that fall within each square bin.
{\it Left to right:} histograms for $L_B/L_T$, $R_e$, and $mag$.
{\it Top to bottom:} histograms for the \gbin{}, \rbin{}, \ibin{}, \zbin{}, and \ybin{} redshift bins.
The lime dashed line $y=x$ in each histogram indicates where the perfect \gampen{} prediction should lie.
Overall, we observe that \gampen{} predictions agree well with the GALFIT results (the best for $R_e$, then for $mag$, then for $L_B/L_T$).
The most noticeable exception is within $L_B/L_T$ histograms of the last two redshift bins, where the linear relationship breaks down---see Section~\ref{sec:perfo_acrc} for a detailed discussion of this and other issues.
\label{fg:pred_true}}
\end{figure}

\begin{figure}
\figurenum{11}
\gridline{\fig{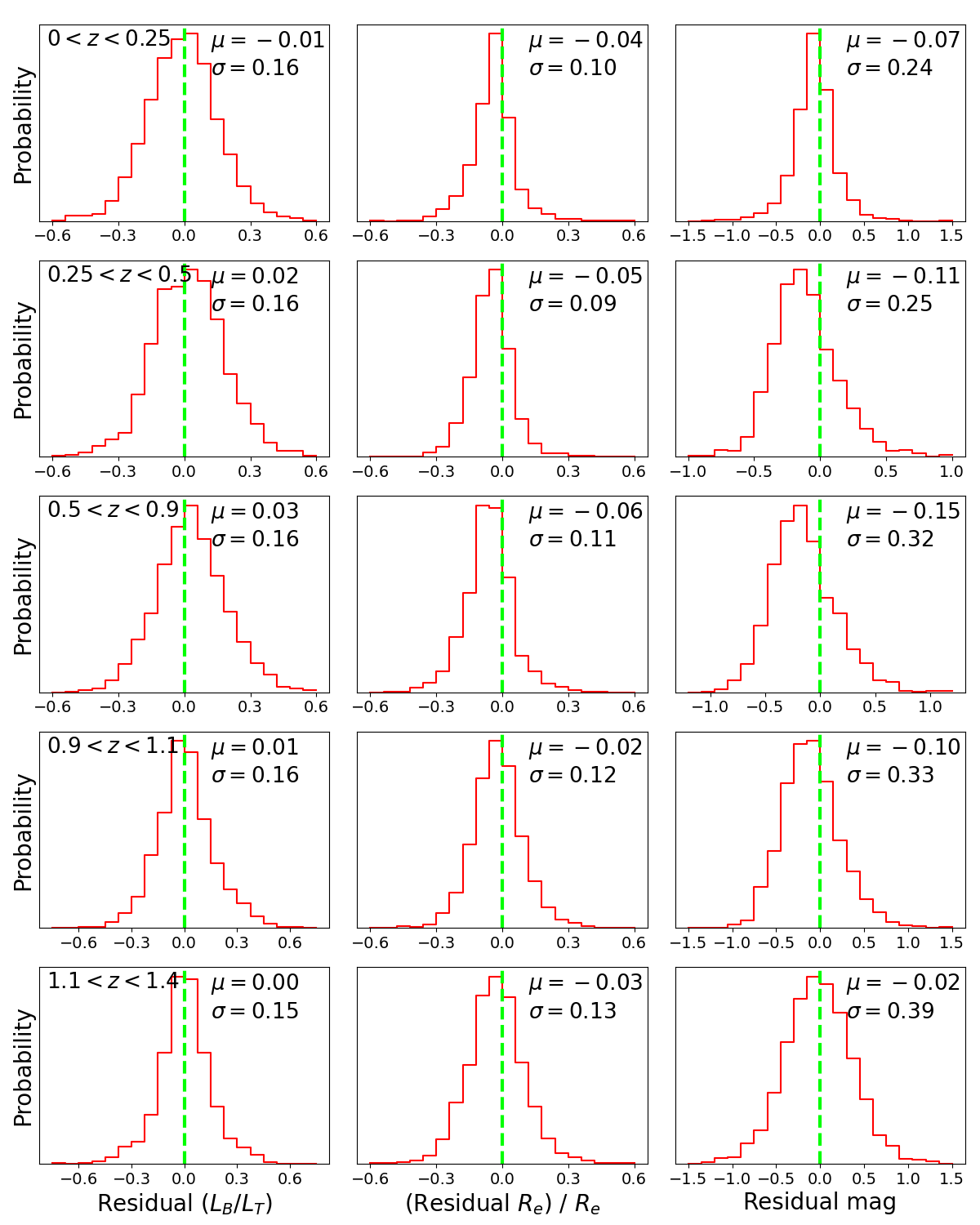}{0.52\textwidth}{}}
\vspace*{-0.4cm}
\caption{
Histograms of residuals (\gampen{} predicted values minus true values from the GALFIT fitting) for galaxies in the common test set, for each of the three morphological parameters, and for each of the five redshift bins.
{\it Left to right:} histograms for $L_B/L_T$, $R_e$, and $mag$.
We use the fractional residual for $R_e$ so that all three x axes are dimensionless.
{\it Top to bottom:} histograms for the \gbin{}, \rbin{}, \ibin{}, \zbin{}, and \ybin{} redshift bins.
Each histogram is normalized to unity, and the lime dashed $x=0$ line marks the perfect case (i.e., zero residual).
In each histogram, we also calculate and attach the mean ($\mu$) and standard deviation ($\sigma$) associated to the distribution.
Briefly summarizing, we find an excellent residual (i.e., a mean close to zero) in $L_B/L_T$ histograms, while the worst residual means are found in $mag$ histograms.
There is also a noticeable increase in the residual standard deviation as we move to higher redshifts in the $mag$ histograms.
See Section~\ref{sec:perfo_acrc}.
\label{fg:hist_rsdl}}
\end{figure}

We now analyze the performance of our models on the entire common test set for each of the five redshift bins.
We start by comparing \gampen{} predicted values (i.e., the most probable values) against ``true'' values determined by GALFIT light profile fitting, for each of the three morphological parameters and for each redshift bin, in order to examine whether our PSFGAN + \gampen{} models can accurately measure these three parameters of the host galaxy using only images of the original galaxy + AGN point source.
Figure \ref{fg:pred_true} shows the corresponding 2D histograms that we make to compare the predicted and true values for different parameters and redshift bins.
We use a common colorbar for all histograms, which is truncated in a way such that square bins with number of galaxies down to 1 are colored yellow (a square bin is colored white if and only if it does not contain any galaxies).
This colorbar is also on logarithmic scale so we can easily highlight bins that contain the majority of galaxies.
In each histogram, we draw a lime dashed line ($y=x$) to indicate where the perfect result should lie (the predicted value equals the true value).
In the majority of cases, the values predicted by our \gampen{} models agree well with the true values from the GALFIT fitting, and the histogram is roughly symmetric with respect to this lime line.
There are, however, two exceptional cases, which we discuss next.

First, for the $L_B/L_T$ histograms in the \zbin{} and \ybin{} redshift bins, the relation between predicted and true values are no longer symmetric about $y=x$---rather, \gampen{} predicts a similar range of $L_B/L_T$ (centered around $L_B/L_T = 0.5$) for both true disk-like and true bulge-like galaxies.
This could be explained by several reasons: 
(1) (PSFGAN) the procedure of adding and removing AGN point sources introduces a pattern at image center that is significant in terms of size and magnitude relative to most host galaxies in the \zbin{} and \ybin{} redshift bins, so there is not much information left to determine the actual $L_B/L_T$ of the host galaxy.
(2) (\gampen{}) the apparent size of the HSC galaxies in the \zbin{} and \ybin{} redshift bins are so small, making it difficult for \gampen{} to do a bulge + disk decomposition.
Recall that decomposition into multiple components is an inherently challenging task when the sizes of the majority of the galaxies become comparable to that of the PSF\footnote{See \citealp{2022ApJ...935..138G}, in which the authors specifically discuss the challenges of doing B+D decomposition for smaller galaxies.}---which is exactly the case in these two bins---although it can still reliably estimate $R_e$ and $mag$. 
Second, in a few $mag$ histograms (especially in the \rbin{}, \ibin{} and \zbin{} redshift bins), \gampen{} predicted values are sightly biased toward smaller magnitude (i.e., larger flux), meaning \gampen{} tends to overestimate the brightness of the host galaxy.
This could be driven if the PSFGAN does not remove the entire added AGN point source (that is, leaving a small positive flux at image center).
We will revisit this phenomenon later in this subsection.

In Figure \ref{fg:hist_rsdl}, we plot histograms of residuals for each morphological parameter and each redshift bin.
Here, we define the residual as predicted value (\gampen{}) minus true value (GALFIT).
Note that for $R_e$ we use the fractional residual (i.e., the residual of $R_e$ divided by the true value of $R_e$) so that the three x axes are dimensionless.
For each histogram, we calculate its mean ($\mu$) and standard deviation ($\sigma$) and attach this information to the subplot along with the lime line (now at $x=0$, still showing the perfect case).
Evidently, the mean of $L_B/L_T$ is very close to zero in all five redshift bins, indicating that there are roughly an equal number of cases of positive and negative residuals.
In comparison, we observe a noticeable deviation of the mean from zero in the $mag$ histograms, which again signifies the fact that our models tend to slightly overestimate host galaxy brightness (which can be explained if a positive residual flux is left after an imperfect AGN point source removal).
Moreover, as we move to higher redshifts, the standard deviation in $mag$ grows remarkably, while it only grows slightly in $R_e$ and remains steady in $L_B/L_T$.

\begin{figure*}[!htb]
\figurenum{12}
\centering
\includegraphics[width=1.0\textwidth]{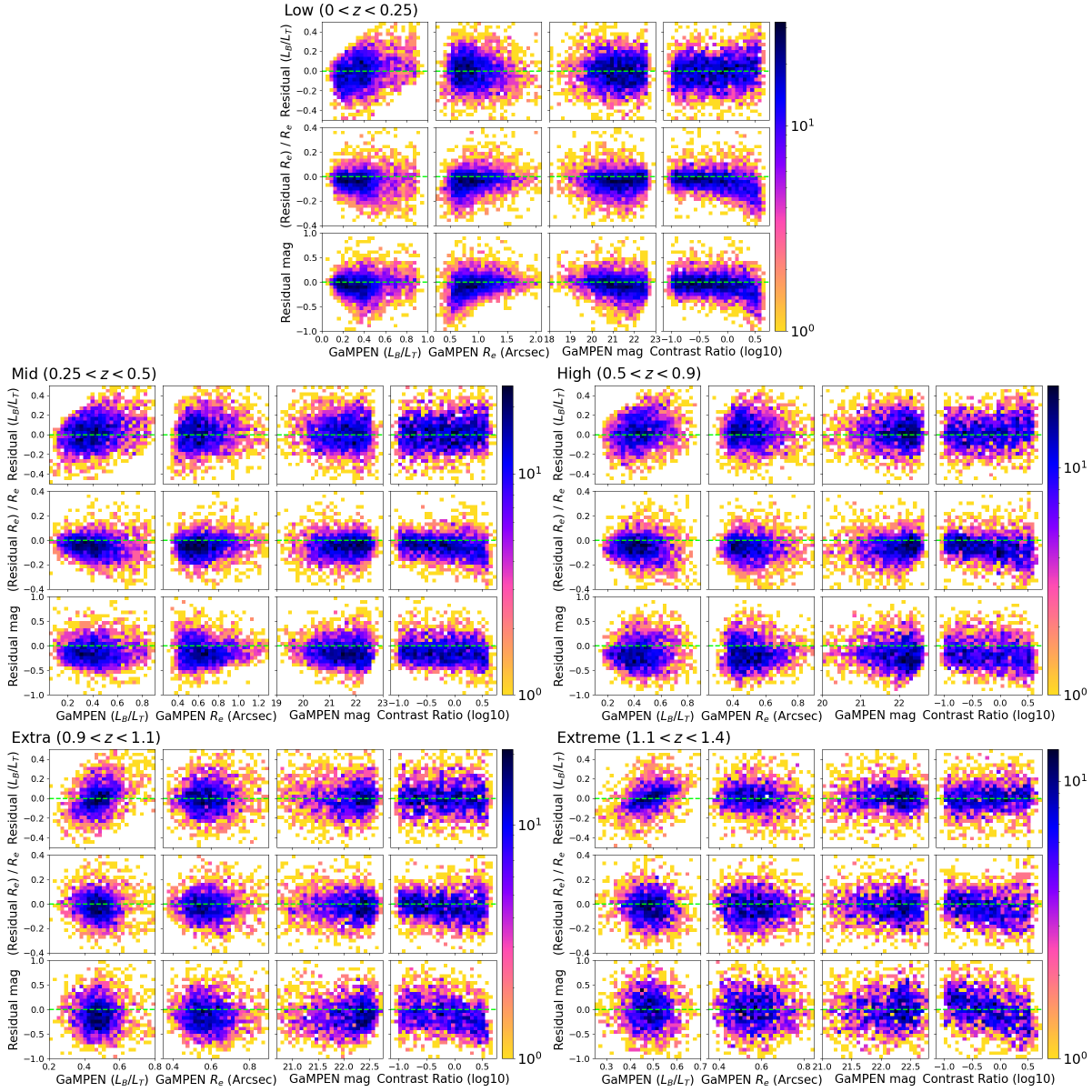}
\caption{
2D histograms of residuals (\gampen{} predicted values minus true values from the GALFIT fitting) vs. \gampen{} predicted values + contrast ratio for galaxies in the common test set, for each of the three morphological parameters and for the \gbin{} (\textit{top panel}), \rbin{} and \ibin{} (\textit{middle panel}) and \zbin{} and \ybin{} (\textit{bottom panel}) redshift bins, respectively.
For each redshift bin, we color histograms using a common logarithmic colorbar to indicate the number of galaxies falling within each square bin.
{\it Left to right:} histograms for \gampen{} predicted $L_B/L_T$, $R_e$, $mag$ and the AGN-to-host-galaxy flux contrast ratio.
{\it Top to bottom:} histograms for residuals of $L_B/L_T$, $R_e$ and $mag$.
We again use fractional residuals for $R_e$ so that the three y axes are dimensionless.
The lime dashed $y=0$ line marks the perfect case (i.e., zero residual).
See Section~\ref{sec:perfo_acrc} for a discussion of all observed trends.
\label{fg:rsdl_pred_all}}
\end{figure*}

Having plotted histograms of residuals, we then explore how these residuals depend on other factors---specifically, how do residuals depend on \gampen{} predicted values of morphological parameters---for if we apply our trained models on real active galaxies in the HSC Wide survey, predicted values (instead of true values) are the only ones we have access to.
In addition to predicted values, we also analyze how residuals depend on AGN-to-host-galaxy flux contrast ratios (recall that we added and then removed an AGN point source for each galaxy). A larger contrast ratio means that we added and removed a brighter AGN point source---thus, the residual flux of the introduced pattern at image center can be larger, leading to a lower accuracy when applying \gampen{} models on the recovered galaxy image.
In Figure \ref{fg:rsdl_pred_all}, we plot 2D histograms of residuals versus \gampen{} predicted values + contrast ratio, for each of the three morphological parameters and each of the five redshift bins.
As in previous plots, the lime line marks the perfect case and a fractional residual for $R_e$, so the three y axes are dimensionless.

In the \gbin{} redshift bin (\textit{top panel} of Figure \ref{fg:rsdl_pred_all}), we find a noticeable dependency: the residuals in $R_e$ and $mag$ are both biased toward the negative end for a smaller and fainter host galaxy and for a larger contrast ratio.
The underlying cause for this is most likely the same as that responsible for the reported biases toward smaller magnitude (larger flux) in Figures \ref{fg:pred_true} and \ref{fg:hist_rsdl}: positive residual flux from imperfect AGN point source removals.
Specifically, a higher contrast ratio can potentially lead to a larger positive residual flux\footnote{As we empirically tested, for bright AGN point sources (contrast ratio greater than $1$), in an overwhelming majority of cases, higher contrast ratios lead to a larger positive residual flux in all five bins. See Appendix \ref{sec:ap:edge_eff} for a discussion of the underlying reason.}, which can certainly add to the total flux of the host galaxy. Thus \gampen{} will likely overestimate host galaxy brightness (i.e., causing a negative residual in $mag$).
Since the size of the introduced pattern (from the imperfect removal of the AGN point source) is usually much smaller than the size of the host galaxy (in the \gbin{} redshift bin), \gampen{} will surely tend to underestimate the host galaxy radius (i.e., causing a negative fractional residual in $R_e$).
By the same reasoning, given a fixed positive residual flux from AGN point source removal, fainter host galaxies are easier to be affected (in other words, the fixed positive residual flux is more significant relative to the flux of a faint host galaxy).
At last, smaller galaxies are statistically fainter in the \gbin{} redshift bin (Figure \ref{fg:pred_true}), therefore, the same phenomenon is again present.

The same observation is repeated in some histograms for the redshift bins \rbin{} and \ibin{} (\textit{middle panel} of Figure \ref{fg:rsdl_pred_all}), although the degree is less severe.
For example, in both bins, it is harder to tell whether smaller and fainter host galaxies will lead to a negative fractional residual in $R_e$. This is because host galaxies are smaller in the \rbin{} and \ibin{} bins than in the \gbin{} bin, so a positive residual flux of the introduced pattern (whose size is close to a PSF) has less effect on affecting the measurement of host galaxy size.
In the last two redshift bins (\textit{bottom panel} of Figure \ref{fg:rsdl_pred_all}), the only tendency (of the same type) that remains is that higher contrast ratios trigger negative residuals in $mag$, again caused by the empirical observation that higher contrast ratios likely lead to positive residual flux from AGN point source removals.
Besides the tendencies in the residuals of $R_e$ and $mag$, there is a noticeable trend in the residual $L_B/L_T$ vs. the predicted $L_B/L_T$ plot in the last two redshift bins---namely, the residual of $L_B/L_T$ becomes larger (smaller) for the higher (lower) predicted $L_B/L_T$.
That is, \gampen{} chooses to assign a $L_B/L_T$ close to $0.5$ for both true disk-like and true bulge-like galaxies in a scenario when the host galaxy flux is substantially affected by imperfect AGN point source removals and there is little information left for \gampen{} to accurately determine the actual $L_B/L_T$ of the host galaxy; the network thus decides to assign $L_B/L_T$ close to $0.5$ so that the overall loss is minimized.

\subsection{Evaluating Model Uncertainties}
\label{sec:perfo_uctt}

\begin{figure*}[!htb]
\figurenum{13}
\centering
\includegraphics[width=1.0\textwidth]{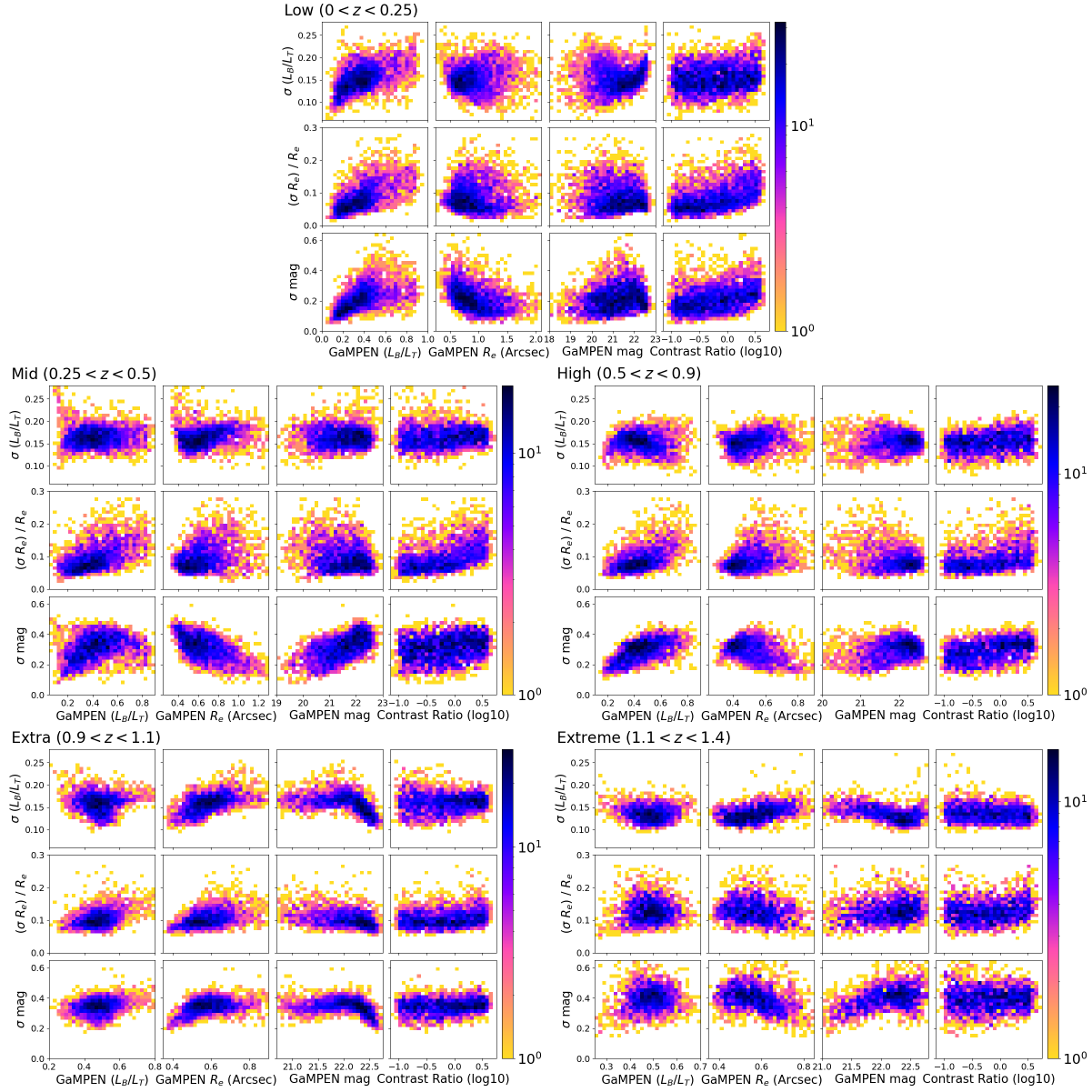}
\caption{
2D histograms of uncertainties (widths of 1$\sigma$ confidence intervals) vs. \gampen{} predicted values and contrast ratio for galaxies in the common test set, for each of the three morphological parameters and for the \gbin{} (\textit{top panel}), \rbin{} and \ibin{} (\textit{middle panel}) and \zbin{} and \ybin{} (\textit{bottom panel}) redshift bins, respectively.
For each redshift bin, we color histograms using a common logarithmic colorbar to better visualize the number of galaxies falling within each square bin.
{\it Left to right:} histograms for \gampen{} predicted $L_B/L_T$, $R_e$, $mag$ and the AGN-to-host-galaxy flux contrast ratio.
{\it Top to bottom:} histograms for uncertainties of $L_B/L_T$, $R_e$, and $mag$.
Just like for residuals, we use fractional uncertainty for $R_e$ so all three y axes are dimensionless.
We report and discuss all the trends in Section~\ref{sec:perfo_uctt}.
\label{fg:sigma_pred_all}}
\end{figure*}

\begin{figure}
\figurenum{14}
\gridline{\fig{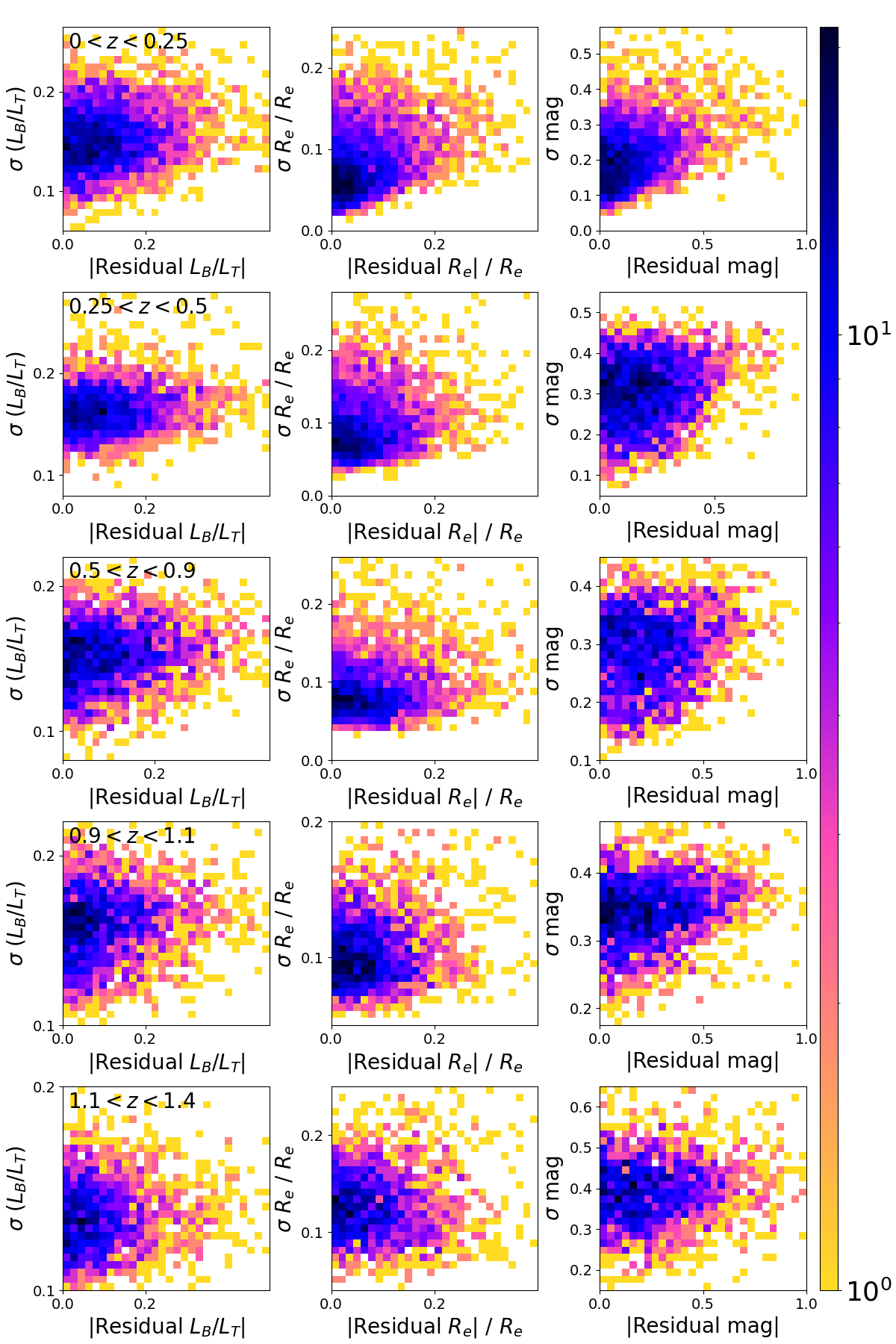}{0.52\textwidth}{}}
\vspace*{-0.4cm}
\caption{
2D histograms of uncertainties vs. residuals for galaxies in the common test set, for each of the three morphological parameters and for each of the five redshift bins.
We color each histogram using a common logarithmic colorbar to better visualize the number of galaxies fall within each square bin.
{\it Left to right:} histograms for $L_B/L_T$, $R_e$, and $mag$.
{\it Top to bottom:} histograms for the \gbin{}, \rbin{}, \ibin{}, \zbin{} and \ybin{} redshift bins.
Like in the previous plots, we use a fractional uncertainty and a fractional residual for $R_e$ so that all the x and y axes are dimensionless.
For residuals, we use their absolute values instead of actual values to demonstrate the relation between the scale of residual and the uncertainty.
See Section~\ref{sec:perfo_uctt} for a detailed discussion.
\label{fg:sigma_rsdl}}
\end{figure}

In addition to the residuals of \gampen {}, it is equally important to explore the uncertainty values predicted by \gampen{} and how this uncertainty depends on the predicted parameters of \gampen{}.
Additionally, as one would reasonably expect, we may also observe higher \gampen{} uncertainties in regions where the \gampen{} residuals are higher.
We analyze distributions related to \gampen{} uncertainties specifically in this subsection.

The \textit{top panel} of Figure \ref{fg:sigma_pred_all} shows 2D histograms of \gampen{} uncertainties versus \gampen{} predicted values + the contrast ratio for the three morphological parameters in the redshift bin \gbin{}.
Here, we define the \gampen{} uncertainty (for each morphological parameter) to be the width of the 1$\sigma$ confidence interval (see Figure \ref{fg:psf_gmp_c}).
Like for the residual of $R_e$, we also use a fractional uncertainty of $R_e$ (equal to the uncertainty of $R_e$ divided by the true value of $R_e$) so that all three y-axes are dimensionless.
In this redshift bin, we observe clear trends in the vast majority of histograms: (1) all three uncertainties are higher for predicted\footnote{Recall that the \gampen{} predicted disk-like (bulge-like) galaxies correlate well with true disk-like (bulge-like) galaxies in the \gbin{} redshift bin (Figure \ref{fg:pred_true}). So the same observation is also valid for true disk-like (bulge-like) galaxies.} bulge-like galaxies than for predicted disk-like galaxies, which can be associated to the fact that in the \gbin{} redshift bin, disk-like galaxies are preferentially larger and are more robust against any residual flux from imperfect AGN point source removals; and (2) all three uncertainties are generally higher for smaller and fainter host galaxies and for larger contrast ratios---a feature that surely makes sense as in these regions it is much harder to determine true parameters of the host galaxy (i.e., the host galaxy flux is easier to be affected by the residual flux from imperfect AGN point source removals).

In the redshift bins \rbin{} and \ibin{} (\textit{middle panel} of Figure \ref{fg:sigma_pred_all}), similar trends are observed, although the details are slightly different.
First, in the majority of cases, the uncertainty of $L_B/L_T$ does not strongly depend on predicted values, despite its dependence on contrast ratio being still present in some cases (e.g., in \ibin{} redshift bin).
Second, the fractional uncertainty of $R_e$ shows a noticeable dependency with the predicted values of the three morphological parameters and the contrast ratio, but its dependencies on the predicted values of $R_e$ and $mag$ are reversed compared to the same histograms in the redshift bin of \gbin{}.
This reveals the fact that in these two bins, \gampen{} is more certain in determining the size of smaller and fainter host galaxies.
Third, dependencies of $mag$ uncertainty remain the same as in the \gbin{} redshift bin.
For the redshift bins \zbin{} and \ybin{} (\textit{bottom panel} of Figure \ref{fg:sigma_pred_all}), the dependency structures are clearly more complex, and in some cases the dependency is no longer monotonic, which indicates that there are potentially multiple uncertainty-residual relations in effect.
(For example, the residual of $L_B/L_T$ vs. predicted $mag$ in the \zbin{} redshift bin shows a clear peak around predicted $mag = 22$, so as the residual of $mag$ vs. predicted $mag$ histogram in the same bin.)
Since the resolution of the HSC Wide survey is limited in the \zbin{} and \ybin{} redshift bins, we detain a detailed discussion of uncertainty dependencies in these two bins in future works, possibly with data release of another survey that has better resolution and signal-to-noise ratios.

Finally, in Figure \ref{fg:sigma_rsdl} we plot uncertainties vs. absolute values of residuals for each of the three morphological parameters and for each of the five redshift bins.
As expected, we observe that in most cases the uncertainty increases with higher residual---which shows that \gampen{} correctly assigns a larger uncertainty in regions where its predictions are less accurate.

\section{Summary and Discussion} \label{sec:summary}
We present a composite ML framework in this paper that is capable of accurately estimating posterior probability distributions of bulge-to-total light ratio ($L_B/L_T$), half-light radius ($R_e$) and flux ($F$) for AGN host galaxies in the Hyper Supreme-Cam Wide survey with $z<1.4$ and $m<23$.
Since morphological parameters can vary greatly with redshift, we divide our training and testing data into five redshift bins: \gbin{} ($0<z<0.25$), \rbin{} ($0.25<z<0.5$), \ibin{} ($0.5<z<0.9$), \zbin{} ($0.9<z<1.1$) and \ybin{} ($1.1<z<1.4$), and in each bin we use images from only one HSC band: \gb{}, \rb{}, \ib{}, \zb{}, and \yb{}, respectively, to make sure we are focusing on a similar range of rest frame wavelengths.
This division makes it possible for us to train separate models in each redshift bin, which can focus on a smaller region of parameters, and therefore, the overall model performance is optimized. 
It also enables us to compare model performance across multiple redshift bins and to reveal possible model performance dependence on redshift.

In each redshift bin, we train a separate model of PSF\-GAN \citep{2018MNRAS.477.2513S} to remove the AGN point source and to recover the host galaxy.
We then train a separate model of \gampen{} \citep{2022ApJ...935..138G} using PSF\-GAN-recovered galaxies.
By empirical testing, we verified the feasibility of training \gampen{} models on PSFGAN-recovered images---namely, \gampen{} predictions on $R_e$ and $F$ are largely unaffected by the presence of AGN point sources at all redshifts, and so as \gampen{} predictions on $L_B/L_T$ in the first three redshift bins ($z<0.9$) (see below and Appendix \ref{sec:ap:edge_eff} for detailed discussion).
Specifically, we use a two-phase approach when training our models.
We first use a large amount of simulated data to train our PSF\-GAN and \gampen{} models from scratch (the ``initial training'' phase).
We then fine-tune the \gampen{} models previously trained on simulated data with respect to labeled real data (the ``transfer learning'' phase).
The major benefit of this approach is that we can use a much smaller set of labeled real data.
The size of simulated data is large, but it is only limited by our computing resources, as one can create an arbitrary number of labeled simulated galaxies.

To create training labels for our real data, we used GALFIT \citep{2002AJ....124..266P} to fit 2D light profiles to $\sim 20,000$ real HSC galaxies in each of the five redshift bins.
These fitted values of $L_B/L_T$, $R_e$ and $F$ are then used properly in the transfer learning phase to fine-tune our \gampen{} models.
We examined comprehensively that predictions from our final PSFGAN and \gampen{} models agree well with these GALFIT-determined truth labels.
In detail, on the common test set (containing a few thousand galaxies each) in each redshift bin, we compared \gampen{} predicted values against GALFIT-determined values (for each of the three morphological parameters).
We find a nice linear correlation (symmetric about $y=x$) between these predicted and true values in $L_B/L_T$ (up to $z=0.9$) and in $R_e$ and $F$ (for the majority of cases in five bins).
This suggests that PSFGAN + \gampen{} works well for $L_B/L_T$ prediction within $z<0.9$, beyond this threshold certain limitations (e.g., signal-to-noise ratio, galaxy radius, etc.) constrains the accuracy for $L_B/L_T$ prediction.
As for $R_e$ and $F$ predictions, PSFGAN + \gampen{} works well till at least $z=1.4$ (and possibly for even higher redshifts), as estimating galaxy size and brightness is intrinsically easier than galaxy morphology.
We explicitly analyzed how \gampen{} residuals (predicted values minus true values) depend on \gampen{} predicted values, which allows us to discard galaxies within particular regions of predicted parameters in which \gampen{} is less accurate or systemically biased.
We also examined \gampen{} predicted uncertainties and verified that their dependencies on \gampen{} predicted values are consistent with empirical observations in the first three redshift bins.

We make our trained PSFGAN and \gampen{} models in all five redshift bins publicly available (Appendix \ref{sec:ap:online_data}).
Using a V100 GPU, the inference time on $\sim 5,000$ galaxies is within $10$ minutes for a trained PSFGAN model.
It takes about a second to process a single feedforward pass for a trained \gampen{} model---if we forward each input image for $1,000$ times to \gampen{}, then an inference on $\sim 5,000$ galaxies (i.e., $5 \times 10^6$ feedforward passes) can be finished within two hours.

This work is a natural extension to our previous works, in which we used PSFGAN with our old ML framework GaMorNet to classify active HSC galaxies by their morphology \citep{2023ApJ...944..124T} or \gampen{} alone to characterize the morphological parameters of inactive HSC galaxies \citep{2022ApJ...935..138G, Ghosh_2023}.
It is also worth mentioning that although we chose to study these three particular parameters ($L_B/L_T$, $R_e$ and $F$), the framework we present in this paper can be adopted to study any other parameters (e.g., axis ratio of the host galaxy).
The same framework can also be easily re-trained via transfer learning with respect to other surveys with different cutout sizes, resolutions, seeing conditions, or signal-to-noise ratios.

We are currently applying our trained PSFGAN and \gampen{} models (present in this paper) on a few catalogs of X-ray confirmed AGN candidates (real active galaxies) in the HSC Wide survey and plan to publish an associated catalog containing details of the morphological parameters of their host galaxies, in order to shed light on the question in-debate of whether bright AGNs are primarily triggered by galaxy mergers.

\section*{Acknowledgments}

The authors sincerely appreciate the anonymous referees for their insightful, encouraging, and pertinent comments, which were extremely helpful in improving the manuscript, clarifying discussions, and making the results more accessible to readers.

This material is based on work supported by the National Science Foundation under Grant No. 1715512.

We acknowledge the support of the National Aeronautics and Space Administration through ADAP grants 80NSSC18K0418 and 80NSSC23K0488. TTA acknowledges support from NASA ADAP grant 80NSSC23K0486.

The authors thank the Yale Center for Research Computing for guidance and use of the research computing infrastructure.

The Hyper Suprime-Cam (HSC) collaboration includes the astronomical communities of Japan and Taiwan, and Princeton University. The HSC instrumentation and software were developed by the National Astronomical Observatory of Japan (NAOJ), the Kavli Institute for the Physics and Mathematics of the Universe (Kavli IPMU), the University of Tokyo, the High Energy Accelerator Research Organization (KEK), the Academia Sinica Institute for Astronomy and Astrophysics in Taiwan (ASIAA), and Princeton University. The funding was contributed by the FIRST program of the Japanese Cabinet Office, the Ministry of Education, Culture, Sports, Science and Technology (MEXT), the Japan Society for the Promotion of Science (JSPS), the Japan Science and Technology Agency (JST), the Toray Science Foundation, NAOJ, Kavli IPMU, KEK, ASIAA, and Princeton University. 

This paper makes use of software developed for the Vera C. Rubin Observatory. We thank the Rubin Observatory for making their code available as free software at http://pipelines.lsst.io/.

This paper is based on data collected at the Subaru Telescope and retrieved from the HSC data archive system, which is operated by the Subaru Telescope and Astronomy Data Center (ADC) at NAOJ. Data analysis was carried out in part with the cooperation of the Center for Computational Astrophysics (CfCA), NAOJ. We are honored and grateful for the opportunity of observing the Universe from Maunakea, which has cultural, historical, and natural significance in Hawaii. 

The Pan-STARRS1 Surveys (PS1) and the PS1 public science archive have been made possible through contributions by the Institute for Astronomy, the University of Hawaii, the Pan-STARRS Project Office, the Max Planck Society and its participating institutes, the Max Planck Institute for Astronomy, Heidelberg, and the Max Planck Institute for Extraterrestrial Physics, Garching, The Johns Hopkins University, Durham University, the University of Edinburgh, Queen's University Belfast, the Harvard-Smithsonian Center for Astrophysics, the Las Cumbres Observatory Global Telescope Network Incorporated, the National Central University of Taiwan, the Space Telescope Science Institute, the National Aeronautics and Space Administration under grant No. NNX08AR22G issued through the Planetary Science Division of the NASA Science Mission Directorate, the National Science Foundation grant No. AST-1238877, the University of Maryland, Eotvos Lorand University (ELTE), the Los Alamos National Laboratory and the Gordon and Betty Moore Foundation.

\clearpage

\appendix

\section{Online Data Access}
\label{sec:ap:online_data}

The source code of our PSFGAN-\gampen{} framework, including instructions of how to effectively use it, is made accessible on GitHub\footnote{\texttt{PSFGAN-\gampen{}} codebase: \url{https://github.com/ufsccosmos/PSFGAN-GaMPEN}.} under a CC-BY-4.0 License and version 1.0.0 is archived in Zenodo \dataset[PSFGAN-GaMPEN]{https://doi.org/10.5281/zenodo.14273617}.

\section{Edge Effects in AGN Point Source Removal} \label{sec:ap:edge_eff}
Like other machine learning models (e.g., \gampen{}), PSFGAN also shows an ``edge effect'' \citep{Ghosh_2023}, meaning parameter ranges where its performance drops noticeably.
Recall that in both the initial training and the transfer learning phases, when adding artificial AGN point sources, we sample the AGN to host-galaxy flux contrast ratio, $R$, from a logarithmic uniform distribution between $-1 < \log R < 0.6$ ($0.1 < R < 3.981$).
Although PSFGAN performs well over the vast majority of this contrast ratio range, its residuals increase when $R$ is very close to one of the two edges.

Specifically, when we apply one of our trained PSF\-GAN models on images of original galaxy + AGN point source whose contrast ratio is around $3.5$ (for instance), only two scenarios are possible: (1) PSFGAN guesses perfectly and removes the entire AGN point source, or (2) PSFGAN guessed contrast ratio is smaller than what it actually is and the network only removes part of the AGN point source flux, leaving a positive residual flux.
The third scenario---in which PSFGAN guessed that the contrast ratio is larger than what it actually is--- can barely happen, since the network itself has not seen many examples of contrast ratios higher than $3.5$ in training.
Therefore, when we combine the only two possible scenarios, the net effect is that PSFGAN tends to leave a positive residual flux around the upper limit of the chosen contrast ratio range.

Likewise, when we are close to the lower limit of the chosen contrast ratio range (e.g., for $R < 0.2$), PSFGAN tends to overestimate the added AGN point source flux and to leave a negative residual flux.
However, around the lower limit of contrast-ratio range, the AGN point source flux is negligible relative to the host galaxy flux, and the negative residual flux of the introduced pattern is less significant as well.
This is distinctively different in comparison to the case when we are close to the upper limit of the contrast ratio range---since we are dealing with very bright AGN point sources, the positive residual flux from an imperfect AGN point source removal can usually be significant relative to the host galaxy.
Combining all these cases, the final effect is that if we explore higher ranges of contrast ratio (say, beyond $R=1$), (from the entire contrast ratio range) there will be more cases of positive, than negative, residual flux that are significant in relative to the host galaxy flux.
These cases of positive residual flux can effectively confuse \gampen{} and make it overestimate host galaxy brightness and underestimate host galaxy size (in cases the host galaxy size is much larger than an averaged size of a PSF).

\section{Fitting Hyper Suprime-Cam Galaxy light profiles with GALFIT} \label{sec:ap:galfitting}
As shown in Section~\ref{sec:hsc_gal}, we prepared cutouts of 20,000 real HSC galaxies in each of our five redshift bins.
In order to use these galaxies in the transfer learning of our PSFGAN and \gampen{} models, we need to acquire reliable morphological parameters for each one of them.
In our previous work \citep{2023ApJ...944..124T}, we only cared about classifying galaxies into three morphological types: {\it disk-dominated}, {\it indeterminate}, and {\it bulge-dominated}.
Thus, we cross-matched catalogs of real HSC galaxies with labeled catalogs from the literature \citep{2011ApJS..196...11S, 2018MNRAS.478.5410D}.
However, in this work, we are exploring a wider redshift range ($0 < z < 1.4$) across five HSC bands.
To the best of our knowledge, there exist no catalog(s) that can provide a uniform, mutually compatible morphology type measurement for a sufficient number of galaxies across all five bins.
Considering we also aim to quantitatively study galaxy's half-light radius (while HSC catalogs only provide Kron radius), we decide to invoke GALFIT, a traditional, widely used algorithm, which fits two-dimensional light profiles to galaxy images for structural parameter determination, to generate reliable morphological parameters for our real HSC galaxies.
In each redshift bin, we use GALFIT to fit the $20,000$ galaxies to determine their bulge-to-total light ratio ($L_B/L_T$), half-light radius ($R_e$), and total flux ($F$).
We use a similar fitting procedure as shown in one of our previous works \citep{Ghosh_2023}, which is presented below.

Before invoking GALFIT, we first run Source Extractor \citep{s_extract} on each input galaxy image to:
\begin{enumerate}
    \item Generate a segmentation map which can be used by GALFIT to mask all secondary objects present in the image.
    \item Make a guess of the morphological parameters for the galaxy to be fitted. These initial guesses will be used as starting values in the light profile fitting step.
    \item Crop the input image\footnote{These cropped images will only be used by GALFIT for light profile fitting. We still use 185-pixel square cutouts with our PSFGAN models. See Section~\ref{sec:prep_trans}.} to a size equal to $10$ $\times$ (source-extractor guessed galaxy half-light radius). If the picked cutout size is larger than the input image size, skip this cropping step for the current input.
\end{enumerate}

\begin{figure*}
\figurenum{A1}
\gridline{\fig{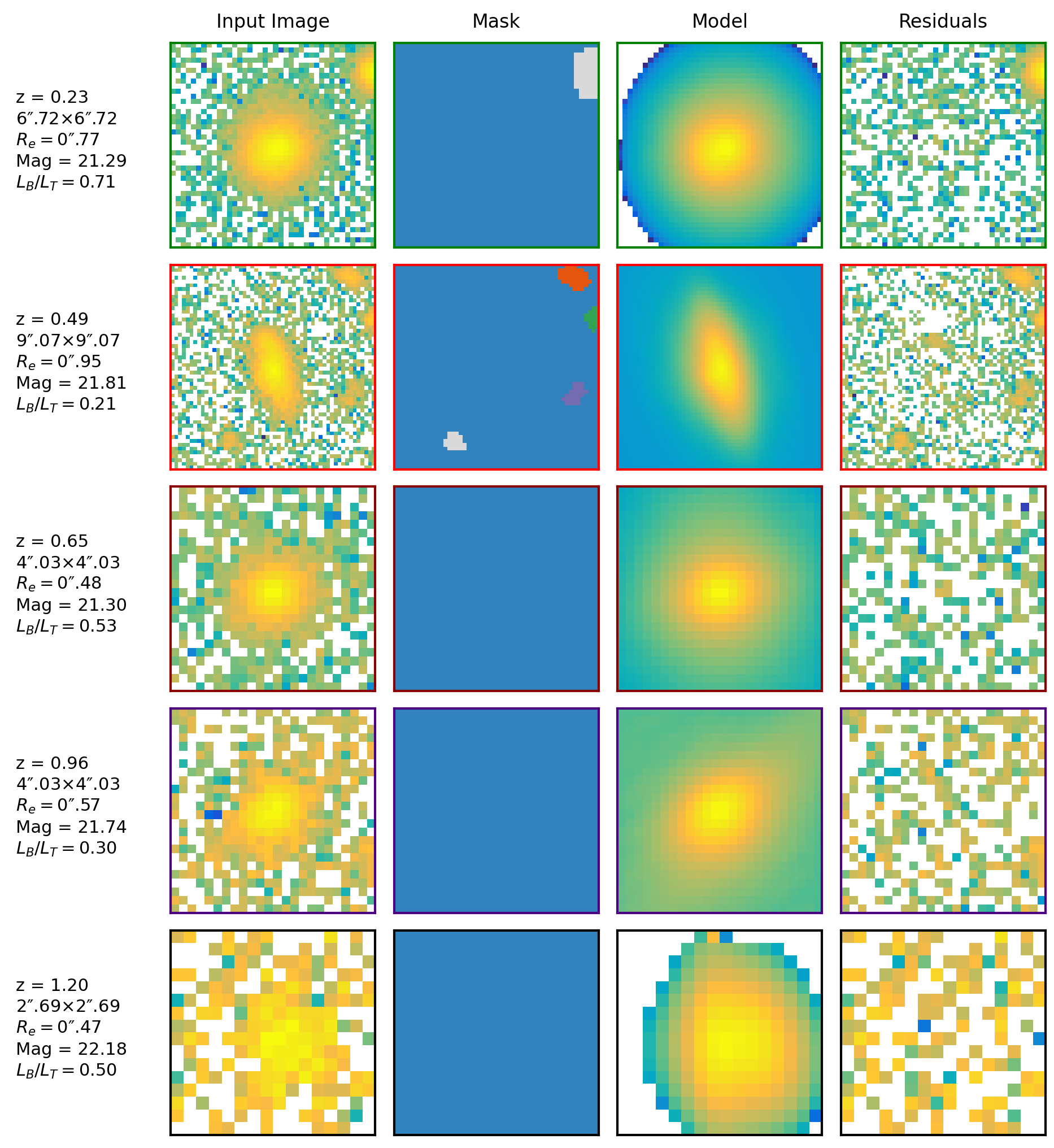}{1.0\textwidth}{}}
\vspace*{-0.6cm}
\caption{An illustration of the light-profile fitting procedure we used with GALFIT. 
{\it Top to bottom}: We show an example for each of the five redshift bins: \gbin, \rbin, \ibin, \zbin, and \ybin.
{\it First column}: The input galaxy image.
{\it Second column}: The mask generated by Source Extractor.
{\it Third column}: The model generated by GALFIT.
{\it Fourth column}: The residual.
{\it Labels on the left}: The photometric redshift, the cutout size picked by Source Extractor, and the GALFIT determined values of half-light radius, magnitude, and bulge-to-total light ratio for each shown example.
\label{fg:galfit_example}}
\end{figure*}

In each of the five bins, we then use GALFIT to fit each galaxy image to a double-component galaxy, which contains a \sersic{} light profile (disk) with a variable index $n$ between $0.8$ and $1.2$ and another \sersic{} light profile (bulge) with a variable index $n$ between $3.5$ and $5.0$.
After entering the source-extractor-guessed parameters as starting values, we run the first round fitting with GALFIT under the following constraints:
\begin{enumerate}
    \item Half-light radius for each of the two components: between $0.5$ and $90$ pixels ($0.084$ and $15.12$ arcseconds)
    \item Magnitude of the galaxy: between $-7.5$ and $7.5$ relative to the Source-Extractor-guessed value
    \item Magnitude difference of the disk and the bulge component: between $-7.5$ and $7.5$ (this is equivalent to having a constraint on the bulge-to-total light ratio ($L_B/L_T$) between $0.001$ and $0.999$)
    \item (Horizontal and vertical) distance between centers of the two components: less than $30$ pixels ($5.04$ arcsecond)
\end{enumerate}
In general, we expect a quick convergence in the first round fitting as we introduced lots of constraints.
Results from the first round fitting are entered as starting values in the second round fitting, which has only the last constraint\footnote{The constraint on \sersic{} indices for both components always in effect.}:
\begin{enumerate}
    \item (Horizontal and vertical) distance between centers of the two components: less than $30$ pixels ($5.04$ arcsecond)
\end{enumerate}
During the second round fitting, GALFIT has an opportunity to explore a wider parameter range, and we expect the convergence takes longer.
We then calculate the half-light radius for the galaxy, using the fitted values of the same parameter for each of the two components, as well as the magnitude of the galaxy.
Figure \ref{fg:galfit_example} shows an example in each of the five bins, including input images, segmentation maps, fitted models, and residuals.
The fitted values of the morphological parameters and the cut-out sizes selected from the source-extractor are shown on the left.

We identify a few cases in which galaxy fits become problematic, in a way similar to our previous work \citep{Ghosh_2023}. To summarize:
\begin{enumerate}
    \item \texttt{non\_converged\_flags} is \texttt{TRUE}. \\
    (GALFIT does not converge)
    \item Calculation of galaxy's half-light radius, based on the half-light radius of the disk and the bulge component, fails.
    \item \texttt{small\_value\_flags} is \texttt{TRUE}. \\
    (Component(s) of fitted galaxy has(have) very small axis-ratio ($< 0.1$) or radius ($< 0.5$ pixels))
    \item \texttt{numerical\_conv\_flags} is \texttt{TRUE}. \\
    (Issues detected with numerical convergence)
    \item \texttt{max\_iters\_flags} is \texttt{TRUE}. \\
    (Max number of iterations (100) reached\footnote{Convergence is not guaranteed in this case.})
    \item Distance between centers ($D_{ct}$) of the two components is too large.
    \item $\chi^{2}$ is too large. (Residual is too large)
\end{enumerate}
Exact upper bounds we used for the distance between centers ($D_{ct}$) and $\chi^{2}$ for each of the five redshift bins are presented in Table \ref{tb:bounds}.

\begin{deluxetable}{cccc}[htbp]
\tablenum{A1}
\tablecaption{Upper Bounds of Distance between Centers ($D_{ct}$) and $\chi^{2}$ in Each Redshift Bin
\label{tb:bounds}}
\tablecolumns{4}
\tablehead{
\colhead{} & \colhead{$D_{ct}$ (pixels)} & \colhead{$D_{ct}$ (arcseconds)} & \colhead{$\chi^{2}$} 
} 
\startdata
    \hline
    \gBin{} ($0 < z < 0.25$) & $12$ & $2.016$ & $2.5$ \\
    \rBin{} ($0.25 < z < 0.5$) & $6$ & $1.008$ & $1.25$\\
    \iBin{} ($0.5 < z < 0.9$) & $4$ & $0.672$ & $1.25$  \\
    \zBin{} ($0.9 < z < 1.1$) & $4$ & $0.672$ & $1.25$ \\
    \yBin{} ($1.1 < z < 1.4$) & $4$ & $0.672$ & $1.25$ \\
    \hline
    \hline
\enddata
\end{deluxetable}

\begin{figure}
\figurenum{A2}
\gridline{\fig{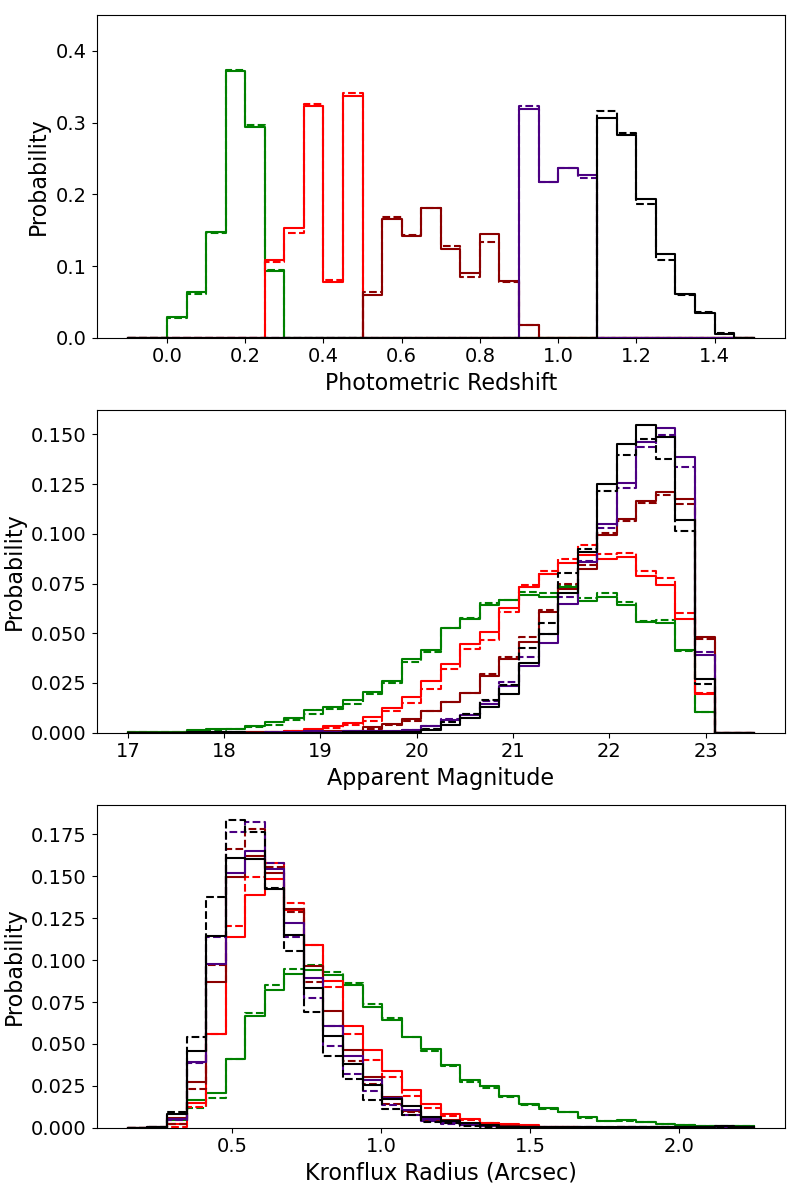}{0.5\textwidth}{}}
\vspace*{-0.4cm}
\caption{Comparison of parameter distributions for pre-GALFIT ({\it solid line}) and post-GALFIT ({\it dashed line}) real HSC galaxies.
{\it Top to bottom}: Comparisons between photometric redshift, magnitude and Kron radius distributions, respectively.
In each panel, we show data from all five redshift bins: \gbin{} ({\it green}), \rbin{} ({\it red}), \ibin{} ({\it dark red}), \zbin{} ({\it purple}) and \ybin{} ({\it black}).
Each histogram is normalized separately to unity.
\label{fg:hsc_gal_para}}
\end{figure}

Under the above constraints and bounds, and starting with $20000$ galaxies in each redshift bin, we arrived at $17789$, $15542$, $13994$, $13706$, and $12519$ galaxies in the \gbin{}, \rbin{}, \ibin{}, \zbin{}, and \ybin{} redshift bins, respectively.
Note that choices of these bounds, as well as some of the constraints, are completely up to the user's preference: the user must reach a sensible balance in either having too few galaxies (rigid constraints) or having too many bad fits (loose constraints), and this is subject to the scientific question at hand.
In Figure \ref{fg:hsc_gal_para} we show distributions in photometric redshift, magnitude, and Kron radius between pre- and post-GALFIT galaxies in all five redshift bins.
It is evident that discarding galaxies from our real HSC galaxy sample in each bin because of bad fits does not selectively exclude galaxies within any specific regions in the parameter space.
These post-GALFIT galaxies (numbers reported above) are the ones we actually use in the transfer learning of our PSFGAN and \gampen{} models (Section~\ref{sec:prep_trans} and Section~\ref{sec:trans_learning}).

\software{PSFGAN \citep{2018MNRAS.477.2513S},
          \gampen{} \citep{2022ApJ...935..138G},
          MLFlow \citep{mlflow},
          TOPCAT \citep{2005ASPC..347...29T},
          GALFIT \citep{2002AJ....124..266P},
          NumPy \citep{2020Natur.585..357H},
          SciPy \citep{2020SciPy-NMeth},
          Astropy \citep{2018AJ....156..123A},
          Pandas \citep{pandas_2010},
          Matplotlib \citep{Hunter:2007}
}

\bibliographystyle{aasjournal}
\bibliography{references}

\end{CJK*}
\end{document}